\documentclass[fleqn,usenatbib]{mnras}

\usepackage{newtxtext,newtxmath}

\usepackage[T1]{fontenc}

\usepackage{soul}

\DeclareRobustCommand{\VAN}[3]{#2}
\let\VANthebibliography\thebibliography
\def\thebibliography{\DeclareRobustCommand{\VAN}[3]{##3}\VANthebibliography}

\usepackage{graphicx}	
\usepackage{amsmath}	
\usepackage{graphicx}
\usepackage{xcolor}
\usepackage{newtxtext,newtxmath}
\usepackage{enumitem}

\newcommand{\sarc}{$^{\prime\prime}$}

\title[Unveiling AGN Outflows]{Unveiling AGN Outflows: [O~{\sc iii}] Outflow Detection Rates and Correlation with Low-Frequency Radio Emission}

\author[Emmy L. Escott]{\parbox{\textwidth}{
Emmy L. Escott,$^{1}$\thanks{E-mail: emily.l.escott@durham.ac.uk}
Leah K. Morabito$^{1,2}$,
 Jan Scholtz$^{3}$,
 Ryan C. Hickox$^{4}$,
 Chris M. Harrison$^{5}$,
 David M. Alexander$^{1}$,
 Marina I. Arnaudova$^{6}$,
 Daniel J. B. Smith$^{6}$,
 Kenneth J. Duncan$^{7}$ ,
 James Petley$^{1,8}$,
 Rohit Kondapally$^{7}$,
 Gabriela Calistro Rivera$^{9}$,
 Sthabile Kolwa$^{10}$\\}
\\
$^{1}$Centre for Extragalactic Astronomy, Department of Physics, Durham
University, South Road, Durham DH1 3LE, UK\\
$^{2}$Institute for Computational Cosmology, Department of Physics, Durham University, South Road, Durham DH1 3LE, UK\\
$^{3}$Kavli Institute for Cosmology, University of Cambridge, Madingley
Road, Cambridge, CB3 OHA, UK \\
$^{4}$Department of Physics and Astronomy, Dartmouth College, 6127 Wilder Laboratory, Hanover, NH 03755, USA \\
$^{5}$School of Mathematics, Statistics and Physics, Newcastle University, NE1 7RU, UK \\
$^{6}$Centre for Astrophysics Research, University of Hertfordshire, College Lane, Hatfield AL10 9AB, UK  \\
$^{7}$Institute for Astronomy, University of Edinburgh Royal Observatory, Blackford Hill, Edinburgh, EH9 3HJ, UK \\
$^{8}$Leiden Observatory, Leiden University, PO Box 9513, 2300 RA Leiden, The Netherlands \\
$^{9}$European Southern Observatory, Karl-Schwarzschild-Straße 2, D-85748 Garching bei M\"{u}nchen, Germany\\
$^{10}$Physics Department, University of Johannesburg, 5 Kingsway Ave, Rossmore, Johannesburg 2092, South Africa\\
}

\date{Accepted XXX. Received YYY; in original form ZZZ}

\pubyear{2024}

\begin{document}
\label{firstpage}
\pagerange{\pageref{firstpage}--\pageref{lastpage}}
\maketitle

\begin{abstract}
Some Active Galactic Nuclei (AGN) host outflows which have the potential to alter the host galaxy's evolution (AGN feedback). These outflows have been linked to enhanced radio emission. Here we investigate the connection between low-frequency radio emission using the International LOFAR Telescope and [O~{\sc iii}] $\lambdaup$5007 ionised gas outflows using the Sloan Digital Sky Survey. Using the LOFAR Two-metre Sky Survey (LoTSS) Deep Fields, we select 198 AGN with optical spectra, 115 of which are detected at 144~MHz, and investigate their low-frequency radio emission properties. The majority of our sample do not show a radio excess when considering radio luminosity - SFR relationship, and are therefore not driven by powerful jets. We extract the [O~{\sc iii}] $\lambdaup$5007 kinematics and remove AGN luminosity dependencies by matching the radio detected and non-detected AGN in $L_{\mathrm{6\muup\\ m}}$ and redshift. Using both spectral fitting and $W_{80}$ measurements, we find radio detected AGN have a higher outflow rate (67.2$\pm$3.4 percent) than the radio non-detected AGN (44.6$\pm$2.7 percent), indicating a connection between ionised outflows and the presence of radio emission. For spectra where there are two components of the [O~{\sc iii}] emission line present, we normalise all spectra by the narrow component and find that the average broad component in radio detected AGN is enhanced compared to the radio non-detected AGN. This could be a sign of higher gas content, which is suggestive of a spatial relationship between [O~{\sc iii}] outflows and radio emission in the form of either low-powered jets or shocks from AGN winds. \par

\end{abstract}




\section{Introduction} \label{introduction}

A significant number of observational studies show that at the centre of almost every galaxy, there is a super-massive black hole \citep[SMBH;][]{kormendy_evidence_1992, kormendy_inward_1995, magorrian_demography_1998} and when a SMBH shows signs of accretion, it is known as an Active Galactic Nucleus (AGN). Some of the most powerful objects in our universe and the most powerful subclass of AGN, which are termed quasars, have the potential to emit a bolometric luminosity of $\mathrm{10^{48}~erg}$ $\mathrm{s^{-1}}$ \citep[see][for a review]{alexander_what_2012}. The gravitational potential of an accreting SMBH can power intense phenomena \citep{antonucci_unified_1993,urry_unified_1995} which can expel vast amounts of gas out of the host galaxy. \par

The activity from the SMBH is believed to have a profound effect on the evolution of the galaxy, a process called AGN feedback, although our only evidence for this is indirect \citep{fabian_observational_2012}. We see in observational data that there is a correlation between the mass of the SMBH and global properties such as the velocity dispersion of the galaxy \citep{gebhardt_relationship_2000,merritt_mbh-sigma_2001}, which suggests the two are linked. Furthermore, cosmological simulations are unable to reproduce the observed universe without including AGN feedback providing heating effects from AGN in their models \citep{bower_breaking_2006,croton_erratum_2006}. AGN feedback can be enacted by powerful phenomena such as shocks induced by winds, ionised and molecular outflows, and jets. However, despite indications that AGN impact how galaxies form and evolve, AGN feedback is still a major open area of research, with many questions waiting to be answered, such as which physical processes are the major cause of turbulent outflows in the interstellar medium (ISM) and how these outflows differ with galactic properties. \par

Processes that generate radio emission are often associated with AGN feedback. One channel is through the most powerful radio jets that commonly originate from AGN which have a much higher radio to optical flux density ratio \citep[radio loud AGN;][]{kellermann_vla_1989}. Powerful radio jets are clearly seen to interact with their environments and host galaxies, for example by driving bulk outflows of ionised gas \citep[e.g.,][]{tadhunter_anisotropic_1989, villar-martin_nature_1999, birzan_systematic_2004}. We see evidence for AGN feedback from these radio jets on large scale ($\sim$100 kpc), e.g from galaxy clusters \citep[e.g.,][]{mcnamara_mechanical_2012, gitti_evidence_2012, timmerman_origin_2022} where jets inflate radio lobes into the ISM producing X-ray cavities. On galactic scales, jets are associated with, for example, disturbed emission line kinematics \citep[e.g.,][]{zakamska_quasar_2014, speranza_multi-phase_2023} or radio lobe expansion \citep{webster_investigating_2021}.

The radio emission from radio-loud AGN is likely due to a radio jet being present \citep{urry_unified_1995}, on the other hand the origin of the radio emission in radio quiet AGN remains unclear. The emission could be produced by low-powered jets, the corona, stellar winds or star formation \citep[see][and references therein]{panessa_origin_2019}. \cite{bonzini_star_2015}, \cite{padovani_vla_2011-1}, and \cite{delvecchio_vla-cosmos_2017} suggest the emission is due to star formation however, other studies \citep[e.g.,][]{white_radio-quiet_2015,macfarlane_radio_2021, calistro_rivera_ubiquitous_2023, yue_novel_2024} propose that the radio emission from radio quiet AGN is due to accretion activity rather than star formation. Moreover, \cite{petley_connecting_2022} connects the radio emission in broad absorption line quasars to the outflow properties, suggesting the radio emission may originate from AGN induced winds.\par

The [O~{\sc iii}] $\lambdaup$5007\ emission line is commonly used as a tracer for warm, ionised gas outflows. These outflows are firmly linked to radio properties, which makes studies of the connection between radio and [O~{\sc iii}] a promising avenue for investigating AGN feedback. \cite{rawlings_relations_1989} shows that the [O~{\sc iii}] and radio luminosities are correlated in a sample of radio galaxies at $z$ $\sim$ 0.5. \cite{mullaney_narrow-line_2013} characterises the [O~{\sc iii}] $\lambdaup$5007\ profiles of 24,264 optically selected AGN at $z$ $\sim$ 0.4 and links the most disturbed [O~{\sc iii}] to being induced by compact radio cores rather than powerful radio jets. Using a sample of 129 uniformly selected radio AGN with $z$~<~0.23, \cite{kukreti_ionised_2023} uses stacking analysis with SDSS spectra to discover that radio AGN with a peak in their radio spectra drives a broader outflowing component in [O~{\sc iii}] than radio AGN without a peak in their spectrum. In the more distant Universe, \cite{nesvadba_sinfoni_2017} presents the [O~{\sc iii}] $\lambdaup$5007 maps and kinematics for 33 $z$~>~2 radio galaxies using VLT/SINFONI imaging spectroscopy, finding complex gas kinematics in all sources and \cite{zakamska_discovery_2016} studies the [O~{\sc iii}] kinematics of red quasars at $z$~$\sim$~2.5 using \texttt{XSHOOTER} on the VLT, discovering very broad and blueshifted [O~{\sc iii}] emission lines. Extending to a sample of radio quiet AGN, \cite{jarvis_prevalence_2019} and \cite{girdhar_quasar_2022} show that jets drive bi-conical [O~{\sc iii}] outflows with the jets strongly interacting at their termini with the ISM. Due to these links between radio emission and [O~{\sc iii}] outflows, after identifying outflows, we can then compare the extracted kinematics to radio data and start to understand how the radio emission is linked to these outflows. \par

The [O~{\sc iii}] $\lambdaup$5007\ emission line is driven by both a gravitational and non-gravitational component. The gravitational component is dependent on the virial motion of the gas of the host galaxy, whereas if an outflow is present, this will produce a non-gravitational component which is commonly seen as the blueshifted shoulder \citep{greene_comparison_2005}. \cite{woo_prevalence_2016} suggests that the enhanced [O~{\sc iii}] velocity dispersion seen with increasing radio luminosity is due to host galaxy properties. They report that the positive relation between [O~{\sc iii}] velocity dispersion and $L_{\mathrm{1.4 GHz}}$ becomes insignificant once normalised by the stellar velocity dispersion. However, more recently \cite{ayubinia_investigating_2023} shows that radio activity does play a role in the enhancement of ionised gas kinematics after normalising in stellar velocity dispersion, suggesting that radio activity yields an additional boost to these outflows. It is still unclear how and to what extent radio emission is linked to the ionised outflows observed in the [O~{\sc iii}] emission line. \par  

The LOw Frequency ARray \citep[LOFAR;][]{haarlem_lofar_2013} has the potential to provide the answer. The LOFAR Two-metre Sky Survey \citep[LoTSS;][]{shimwell_lofar_2017, shimwell_lofar_2019, shimwell_lofar_2022} is an ongoing survey, which aims to cover the entire northern hemisphere at a frequency of 144~MHz. With this data, 6\sarc\ resolution imaging is possible with an average sensitivity of $\sim$70$\muup$Jy/beam. Due to LOFAR's wide field of view and high level of sensitivity, it is a powerful instrument to conduct deep field surveys. Alongside LoTSS there is also a data release for three deep fields: Lockman Hole, European Large Area Infrared Space Observatory Survey-North 1 (ELAIS-N1) and Bo\"{o}tes \citep{sabater_lofar_2021, tasse_lofar_2021}. The first deep fields data release reached sensitivities of 22, 20, and 30$\muup$Jy/beam, respectively, which before LoTSS was unprecedented at such low frequencies. To emphasise how deep this survey is, if LOFAR operated at the same $L_{\mathrm{1.4GHz}}$ frequency as the VLA (Very Large Array), then Bo\"{o}tes and Lockman would have sensitivities (assuming a typical synchrotron spectral index of -0.7) of 6$\muup$Jy/beam and 5$\muup$Jy/beam, respectively. Comparatively the VLA-COSMOS Deep Survey \citep{schinnerer_vla-cosmos_2004, schinnerer_vla-cosmos_2007} reaches sensitives between 7-15$\muup$Jy/beam at this frequency, but only covers $\sim$ $\mathrm{2deg^{2}}$ of the sky, whereas the LoTSS Deep Fields combined cover a much larger field of view of $\sim$ $\mathrm{27deg^{2}}$, providing bigger samples for statistical studies. With such deep data at low frequencies we can probe significantly faint radio emission produced predominately by synchrotron emission, without susceptibility to free-free contamination, to further our understanding of the radio sky and large samples of low radio luminosity AGN can be obtained and studied.\par

This is the first of two papers to help us further understand the link between [O~{\sc iii}] and low frequency radio emission. Within this first paper, we fit the [O~{\sc iii}] emission line to investigate [O~{\sc iii}] kinematics of AGN within the LoTSS Deep Fields to investigate the link between radio emission and [O~{\sc iii}] kinematics for a well-defined sample of AGN. This allows us to compare the [O~{\sc iii}] properties between AGN detected within the LoTSS Deep Fields and AGN that are not detected. For the second paper, we will fully utilise the capabilities of LOFAR by studying the radio morphologies of these AGN at sub-arcsecond resolution using images from \cite{sweijen_deep_2022}, \cite{de_jong_into_2024}, and Escott et al. in prep. This will allow us to understand the radio emission of these AGN down to sub-galactic scales and link these morphologies to [O~{\sc iii}] kinematics and outflows. \par 

In this paper, we first summarise the data and sample selection in Section \ref{data}. In Section \ref{methods} we outline the spectral fitting procedure and treatment of [O~{\sc iii}] outflows. We present the results of this paper in Section \ref{results} and then the discussion and conclusions are in Section \ref{discussion} and \ref{conclusion} respectively. In this work we assume a WMAP9 cosmology \citep{hinshaw_nine-year_2013} with $H_{0}$ = 69.32~km~$\mathrm{s^{-1}}$ $\mathrm{Mpc^{-1}}$, $\Omega_{m}$ = 0.287, and $\Omega_{\Lambda}$ = 0.713. \par

\section{Data} \label{data}

\subsection{LoTSS Deep Fields} \label{Deep Fields}

We focus on all three LoTSS Deep Fields: ELAIS-N1, Lockman Hole, and Bo\"{o}tes \citep{sabater_lofar_2021,tasse_lofar_2021} ($\sim$7.15deg$^2$, $\sim$10.7deg$^2$, and $\sim$9.5deg$^2$ respectively). We use the Deep Fields in LoTSS due to Lockman Hole and ELAIS-N1 benefiting from high-resolution images presented in \citep{sweijen_deep_2022} and \citep{de_jong_into_2024}. A similar image will soon be available for the Bo\"{o}tes field (Escott et al. in prep). In a follow-up paper we will use these high resolution images to further understand the link between [O~{\sc iii}] and radio emission. Furthermore, we use these Deep Fields because the low-frequency radio emission is $\sim$4 times deeper than the standard DR2 LoTSS images meaning we can probe a new parameter space, down to fainter radio luminosities. These fields also provide excellent ancillary data. The observations are at a central frequency of 144~MHz for both Bo\"{o}tes and Lockman Hole, and 146 MHz for ELAIS-N1, with a total integration time of 80 hours for Bo\"{o}tes, 112 hours for Lockman Hole and 164 for ELAIS-N1. Bo\"{o}tes has a central sensitivity of 32$\muup$Jy, for Lockman Hole the central sensitivity is 22$\muup$Jy, and 20$\muup$Jy for ELAIS-N1, with the noise increasing towards the edge of the images and bright sources.\par

\subsubsection{Multi-Wavelength Data} \label{muliti-data}

The multi-wavelength catalogue for the LoTSS Deep Fields is described in \cite{kondapally_lofar_2021} but we summarise the data here. The LoTSS Bo\"{o}tes Deep Field catalogue is cross-matched with the NOAO Deep Wide Field Survey \citep[NDWFS;][]{jannuzi_noao_1999} which are detected in the I and IRAC 4.5$\muup$m bands \citep{brown_evolving_2007, brown_red_2008} where \cite{brown_evolving_2007} combines 15 multi-wavelength bands between 0.14$\muup$m and 24$\muup$m from various surveys with PSF matched aperture photometry. The catalogue also contains: Ultra Violet (UV) from GALEX surveys, Near Infrared (NIR) from \cite{gonzalez_newfirm_2010}, as well as Mid Infrared (MIR) from the $Spitzer$ Deep Wide Field Survey \citep[SDWFS;][]{ashby_Spitzer_2009}, and Far Infrared (FIR) data from HerMES and MIPS. \par

The multi-wavelength catalogue for Lockman Hole has the largest area of the Deep Fields covered in \cite{kondapally_lofar_2021} and has wavelength coverage between 0.15$\muup$m and 500$\muup$m. The MIR data uses the IRAC instrument on the $Spitzer$ Space Telescope \citep{werner_Spitzer_2004}, more specifically, SWIRE \citep{lonsdale_swire_2003} and the $Spitzer$ Extragalactic Representative Volume Survey \citep[SERVS;][]{mauduit_Spitzer_2012}. For more details on both fields consult \cite{kondapally_lofar_2021}. \par

ELAIS-N1 has the same wavelength coverage as Lockman Hole, 0.15-500$\muup$m, and the MIR data also uses SWIRE AND SERVS. The UV data is from the Deep Imaging Survey (DIS) using GALEX \citep{martin_galaxy_2005, morrissey_calibration_2007}, the NIR data is from the UK Infrared Deep Sky Survey (UKIDISS) Deep Extragalactic Survey (DXS) DR10 \citep{lawrence_ukirt_2007}, and FIR from HerMES and MIPS. The optical data comes from the Pan-STARRS-1 \citep{kaiser_pan-starrs_2010} and HSC-SSP \citep{aihara_first_2018}.

We use derived properties for the radio detected AGN such as star formation rates (SFR) and stellar masses from the catalogue described in \cite{best_lofar_2023}. The authors use Spectral Energy Distribution (SED) fitting to calculate these derived properties. These SEDs span from the UV to FIR and from these the authors obtain a consensus SFR and consensus stellar mass. When an AGN component is present, this is accounted for within the SED fitting code. \par 

\subsection{Optical SDSS Spectra}

In this work, we use optical spectra from the Sloan Digital Sky Survey (SDSS). To identify these spectra, we use two catalouges: the SDSS DR16 Quasar catalogue \citep{lyke_sloan_2020} and a broad-line AGN catalogue from \citep{liu_comprehensive_2019}. The DR16 Quasar catalogue \citep{lyke_sloan_2020} consists of 750,414 spectroscopically confirmed quasars across 9,376 square degrees. The broad-line AGN catalogue, \cite{liu_comprehensive_2019}, is composed of 14,584 AGN at $z$ < 0.35, identified by the width of Balmer emission lines, particularly H$\alpha$. \cite{lyke_sloan_2020} includes all quasar sources classified by SDSS I,II and III, while the catalogue from \cite{liu_comprehensive_2019} identifies broad-line AGN among SDSS DR7 sources. The SDSS spectra associated with these catalogues are from the original SDSS spectrograph or the newer BOSS instrument. The wavelength coverage for the original spectrograph is 3800-9200~\AA, compared with the improved BOSS coverage of 3650-10400~\AA\ with a spectroscopic resolution of 1500 at 3800~\AA, and 2500 at 9000~\AA\ . These catalogues together fill the redshift-luminosity space with AGN at lower redshifts \citep{liu_comprehensive_2019} and quasars at higher redshifts \citep{lyke_sloan_2020}. We use these two catalogues as the basis to select our sample as described in Section \ref{Optical}.\par

\subsection{Sample Selection} \label{Optical}

To construct our sample, we first remove any sources in \cite{liu_comprehensive_2019} that are duplicated within \cite{lyke_sloan_2020} and then apply Multi-Order Coverage Maps (MOCs) of the respective Deep Fields \citep{kondapally_lofar_2021} so that radio data from the LoTSS Deep Fields (see Section \ref{Deep Fields}) is available. Bright sources may also affect the data quality of some areas of the sky. We therefore apply starmasks \footnote{The MOCs and starmasks are available at \url{https://lofar-surveys.org/deepfields.html}} to each field to remove this unreliable data. \par

The next step is to ensure [O~{\sc iii}] spectral information is accessible so we can analyse the emission line's kinematics. We remove AGN from our sample with $z$ > 0.83, to ensure we can analyse [O~{\sc iii}] equally across the AGN. This leaves 332 AGN remaining. \par

The spectra for seven sources could not be downloaded from the SDSS spectral search and upon manually searching their plate details it appeared these spectra had no spectral lines which indicates potential incorrect classification, hence we remove these from our sample. We remove an additional two AGN in Bo\"{o}tes. One is removed because upon visual inspection we find that [O~{\sc iii}] and H$\beta$ are double peaked emission lines, which can be characteristic of dual AGN \citep{zhou_obscured_2004}. Very Long Baseline Interferometry (VLBI) imaging at 1.7~GHz and 5~GHz from \cite{frey_two_2012} further supports that this object is a dual AGN. For the other source, we choose to discard this due to lack of emission lines in the optical spectrum, including [O~{\sc iii}]. It is possible that this is a mis-identification or this AGN could be an optically quiescent quasar \citep{greenwell_candidate_2021}, which maybe obscured due to a `cocooned' phase of the NLR (Narrow Line Region). To confirm this, we would require the infrared (IR) spectrum to see if there is strong [O~{\sc iii}] emission present and an X-ray observation to see if the emission can be identified as an AGN. In the Lockman Hole field, we remove three AGN after attempting to fit the spectra. It became clear there is not enough information in the spectral window to provide robust fits and therefore we remove the sources from the sample. This leaves a sample of 320 AGN, 100 in Bo\"{o}tes, 137 in Lockman Hole and 83 in ELAIS-N1. \par

The results of this paper rely on extracting information from spectra, so we implement a Signal-to-Noise (SNR) cut of 5, where the signal is taken from the continuum subtracted [O~{\sc iii}] $\lambdaup$5007, while the noise is measured from two line-free spectral windows at 5040-5060~\AA\ and 4760-4840~\AA\ . We also test SNR cuts of 3 and 10 and find the results remain consistent, therefore we use a cut of 5 as a balance between high SNR and sample size. The mean SNR of the 320 AGN is 15.32 and using a cut of 5 we reduce the sample from 320 AGN to 217 AGN with a similar percentage of sources removed in each field. The SNR cut is necessary because, as discussed in Section \ref{OIII fitting}, distinguishing between broad, weak outflows and noise can be difficult for low SNR sources. \par

To further ensure that any [O~{\sc iii}] emission is real and not noise, we incorporate a cut which removes AGN which have a very broad narrow component fitted to [O~{\sc iii}], where the peak of this narrow component is not significantly higher than the RMS (root mean squared) noise. If an AGN has a single component with a FWHM (Full Width Half Max) greater than 850~km~$\mathrm{s^{-1}}$ and the peak of the fitted component is less than 5 times the RMS of [O~{\sc iii}], we remove these sources. This leaves 208 AGN. \par   

We cross-match all 208 AGN with the optical and IR catalogue from \cite{kondapally_lofar_2021} using a search radius of 2\sarc\ . We then ensure that MIR data are available for all of these sources which we can use as a proxy for AGN luminosity as described in Section \ref{bias}. We discover that two sources (105057.30+593214.4 and 105902.18+584008.2) do not contain channel 3 and/or channel 4 $Spitzer$ information and therefore we are unable to conduct the matching process described in Section \ref{bias} and hence we remove these two sources from our sample. This leaves 206 AGN. \par

As part of the SDSS targeting procedure, in DR16Q many AGN are targeted because they are detected within 1 \sarc\ of a FIRST radio source, which could provide a radio-biased sample. We therefore remove 8 sources with either a \texttt{QSO\_EBOSS\_FIRST}, \texttt{QSO1\_EBOSS\_FIRST} or \texttt{QSO\_FIRST\_BOSS} flag, all of which are located in Bo\"{o}tes. \par

This leaves us with a final sample of 198 AGN. 90 reside in Lockman, 55 in Bo\"{o}tes, and 53 in ELAIS-N1.

\subsubsection{Radio Properties of the Sample} \label{radio data}

The multi-wavelength catalogue from \cite{kondapally_lofar_2021} has already been cross-matched with the LoTSS data, providing radio information. As mentioned previously, the flux limits for the three Deep Fields are different. There is only one 5$\sigmaup$ source in Lockman Hole and another in ELAIS-N1 which is below the flux limit in the Bo\"{o}tes field. We therefore move these sources (ILTJ104811.63+591047.6 and ILTJ160946.89+550533.3) from the radio detected to radio non-detected population to provide an effectively consistent flux limit for the sample, but note that this does not change the results. We find that 115 sources have a LoTSS detection: 33 in Bo\"{o}tes, 55 in Lockman Hole, and 27 in ELAIS-N1. This leaves 83 AGN without a detection; 22 in Bo\"{o}tes, 26 in ELAIS-N1, and 35 in Lockman Hole. \par

We calculate the 144~MHz $k$-corrected luminosity using the spectral index, $\alpha$ (where $S_{\nu} \propto \nu^{-\alpha}$), assuming a typical synchrotron spectral index of $\alpha=0.7$ \citep{klein_radio_2018} and use the spectroscopic redshift from SDSS. For the radio non-detected AGN, we present upper limits calculated using the median value of an aperture with radius 3\sarc\ from the RMS map of the relevant field \footnote{RMS maps are available at \url{https://lofar-surveys.org/deepfields.html}} to ensure any possible emission is captured. \par

The final catalogue of the 198 AGN will be available on CDS upon publication, containing both optical and radio data, the fitting results described in Section \ref{OIII fitting} will be available upon request. \par

\subsection{$L_{6\mu m}$-$z$ Matching} \label{bias}

\begin{figure}
    \centering
    \includegraphics[width=0.5\textwidth]{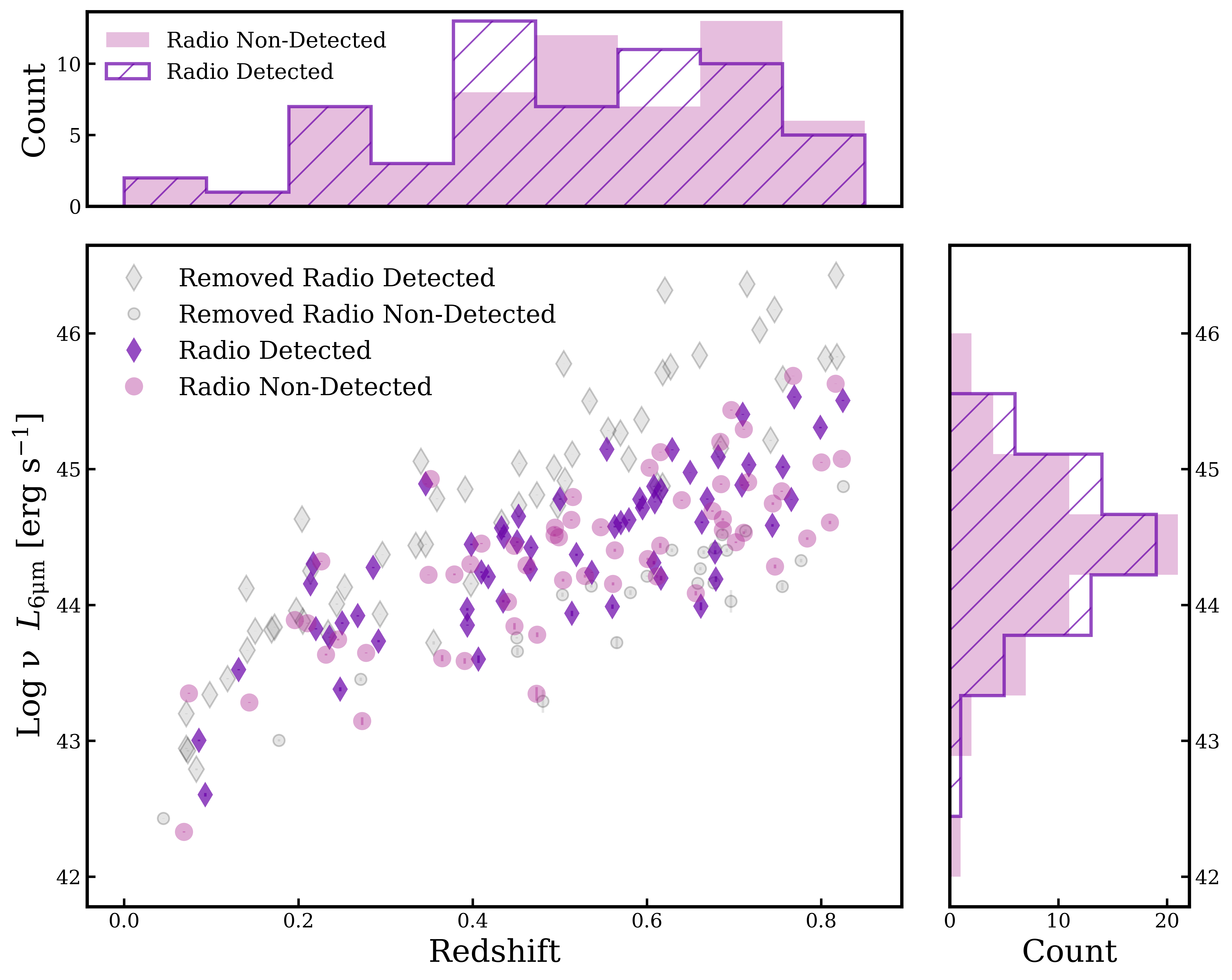}
    \caption{The $L_{\mathrm{6\muup\\ m}}$ and redshift relation of our radio and radio non-detected populations. The coloured markers show one iteration from the 1000 trials to match a radio detected AGN (purple diamonds) in $L_{\mathrm{6\muup\\ m}}$ and redshift to a radio non-detected AGN (pink circles). The grey points represent the AGN which, in this iteration, are removed as these are unmatched, with diamonds portraying the radio detected AGN, and circles for the radio non-detected AGN. The top histogram conveys the redshift distribution and the right histogram is the distribution of $L_{\mathrm{6\muup\\ m}}$. The hashed purple histogram is the radio detected AGN and the pink solid histogram are the radio non-detected AGN, showing they present similar distributions.}
    \label{fig:matched}
\end{figure}

Differences seen in [O~{\sc iii}] between our radio detected and non-detected AGN could be driven by a luminosity bias. The MIR luminosity at 6$\muup$m, $L_{\mathrm{6\muup\\ m}}$, is a popular tracer for AGN bolometric luminosity \citep{richards_spectral_2006} because UV emission produced by the central AGN is absorbed by dust from the torus or NLR and gets re-emitted with MIR wavelengths, providing an orientation-free proxy for the AGN luminosity. Therefore $L_{\mathrm{6\muup\\ m}}$ can be used to remove an intrinsic luminosity bias between populations. We adopt a common method to match the populations in $L_{\mathrm{6\muup\\ m}}$ and redshift \citep[e.g.,][]{rosario_fundamental_2020, andonie_panchromatic_2022, fawcett_striking_2023}. We calculate $L_{\mathrm{6\muup\\ m}}$ for AGN within Bo\"{o}tes using the interpolated aperture and extinction corrected flux densities from channel 3 (5.8$\muup$m) and channel 4 (8$\muup$m) of the $Spitzer$ Deep, Wide-Field Survey \citep[SDWFS;][]{ashby_Spitzer_2009}, and for Lockman Hole, as well as ELAIS-N1, we use the $Spitzer$ Wide-Area Infrared Extragalactic Survey \citep[SWIRE;][]{lonsdale_swire_2003} fluxes, which are both included in \cite{kondapally_lofar_2021}. To begin the matching process we start with the radio non-detected AGN (as this has the smaller population), and then we randomly select from the radio detected population so that each radio non-detected AGN has an equivalent radio detected AGN. We set the tolerance to be $\Delta$ $z=$ 0.06 and $\Delta$ log $L_{\mathrm{6\muup\\ m}}$ = 0.3. \par

In Figure \ref{fig:matched} we see both the total sample without any matching alongside the closest matched population, where starting from the radio non-detected AGN, the closest radio detected AGN has been matched to said radio non-detected AGN. This closest matched population contains 118 AGN in total, 59 radio detected and 59 radio non-detected and are indicated by coloured markers in Figure \ref{fig:matched}. When considering all 198 AGN, We can see that the average $L_{\mathrm{6\muup\\ m}}$ is higher in the radio detected AGN. However, after matching, the distribution of $L_{\mathrm{6\muup\\ m}}$ and redshift is now similar for the two populations. To statistically quantify the differences between the populations from the closest matched population portrayed in Figure \ref{fig:matched}, we perform a 2D Kolmogorov-Smirnov (KS) test after the populations are matched between the $L_{\mathrm{6\muup\\ m}}$ and redshift of the two populations using the public code \texttt{ndtest}\footnote{Written by Zhaozhou Li, \url{https://github.com/syrte/ndtest}} and obtain a p-value of 0.61, confirming that the two populations are statistically indistinguishable.\par

\section{[O~{\sc iii}] emission line profiles and outflow identification} \label{methods}
\subsection{Spectral Fitting} \label{OIII fitting}

To model the emission line profile of the [O~{\sc iii}] emission line and identify outflows, we use the fitting module of \texttt{QubeSpec} IFU analysis code \footnote{Available at \url{https://github.com/honzascholtz/Qubespec}} \citep{scholtz_impact_2021}. We simultaneously fit the [O~{\sc iii}] $\lambdaup\lambdaup$4959,5007 and H$\beta$ $\lambdaup$4861 emission lines and their neighbouring continuum between $\lambdaup$4800~\AA \space and $\lambdaup$5100~\AA \space. To describe emission line profiles, each line is modelled with one or two Gaussian components, with the line centroids, FWHM and fluxes (normalisation) as free parameters. In each case the continuum is well characterised by a power-law with slope and normalisation as a free parameters. For the [O~{\sc iii}] $\lambdaup\lambdaup$4959,5007~\AA \space emission line doublet we simultaneously fit [O~{\sc iii}] $\lambdaup$4959~\AA \space and [O~{\sc iii}] $\lambdaup$5007~\AA, using the respective vacuum rest-frame wavelengths of 4960.3~\AA \space and 5008.24~\AA. For both the narrow component and broad component (where present), we tie the line widths and central velocities of the two lines and fix the [O~{\sc iii}] $\lambdaup$5007/[O~{\sc iii}] $\lambdaup$4959 flux ratio to be 2.99 \citep{dimitrijevic_flux_2007}. We initially fit the continuum, [O~{\sc iii}], and H$\beta$ models. \par

As some of our targets are luminous Type-1 AGN, we need to take into consideration the Fe~II emission, originating from close to the accretion disk. This emission originates from a series of blended faint Fe~II transitions that can appear as subtle features in the continuum that can mimic faint [O~{\sc iii}] broad components. If we detect a broad H$\beta$ component indicating a BLR (Broad Line Region), we refit the spectra with all the models including the Fe II templates. We follow the approach of \cite{kakkad_super_2020} and use different templates: no Fe~II template, Veron, BG92, and Tsuzuki. We take the Veron template from the works of \cite{veron-cetty_unusual_2004}, which takes the spectrum of I Zw~1 and uses the intensities of the broad and narrow Fe~II lines to create a Fe~II template. BG92, \cite{boroson_emission-line_1992}, removes all emission lines from I Zw~1 that are not Fe~II and creates a template from this information. Finally, Tsuzuki from \cite{tsuzuki_fe_2006} obtains further spectral information for I Zw~1 and separates Fe~II emission from Mg~II $\lambdaup$2798 in order to create this template. In each case, these templates are fitted after convolving with a Gaussian profile with a width of 2000-6000~km~$\mathrm{s^{-1}}$ \citep{park_new_2022}. The free parameters for each template are Gaussian convolution width and flux normalisation of the template. \par

We do not couple the width and redshifts of H$\beta$ to the [O~{\sc iii}] as these emission lines can have different line profiles in luminous AGN such as the ones we are investigating in this work \citep{scholtz_kashz_2020,scholtz_impact_2021}. Furthermore, our observations are not sensitive enough to detect the faint outflow component in H$\beta$. Instead, two Gaussian profiles describe the emission from the NLR and BLR regions (if present). The H$\beta$ and [O~{\sc iii}] velocity offset range from -180~km~$\mathrm{s^{-1}}$ to 193~km~$\mathrm{s^{-1}}$ for our matched AGN. \par 

We use the Bayesian Information Criterion (BIC) to assess the quality of the fit and select the model with the lowest BIC score as best representing the data. We find the best-fit parameters using a Markov Chain Monte Carlo \citep[MCMC;][]{goodman_ensemble_2010} method, implemented using the \texttt{emcee} python library \citep{foreman-mackey_emcee_2013}. We set uniform priors in log space for the normalisations of the continuum, Gaussian profiles and Fe~II templates, for example the FWHM of the broad component of [O~{\sc iii}] was set to 2000~km~$\mathrm{s^{-1}}$ and allowed to deviate between 500~km~$\mathrm{s^{-1}}$ above and below this value. The final parameters quoted in this paper are the $\mathrm{50^{th}}$ percentile of the posterior distribution with errors describing the 68 percent confidence interval. \par 

\begin{figure*}
    \centering
    \includegraphics[width=\textwidth]{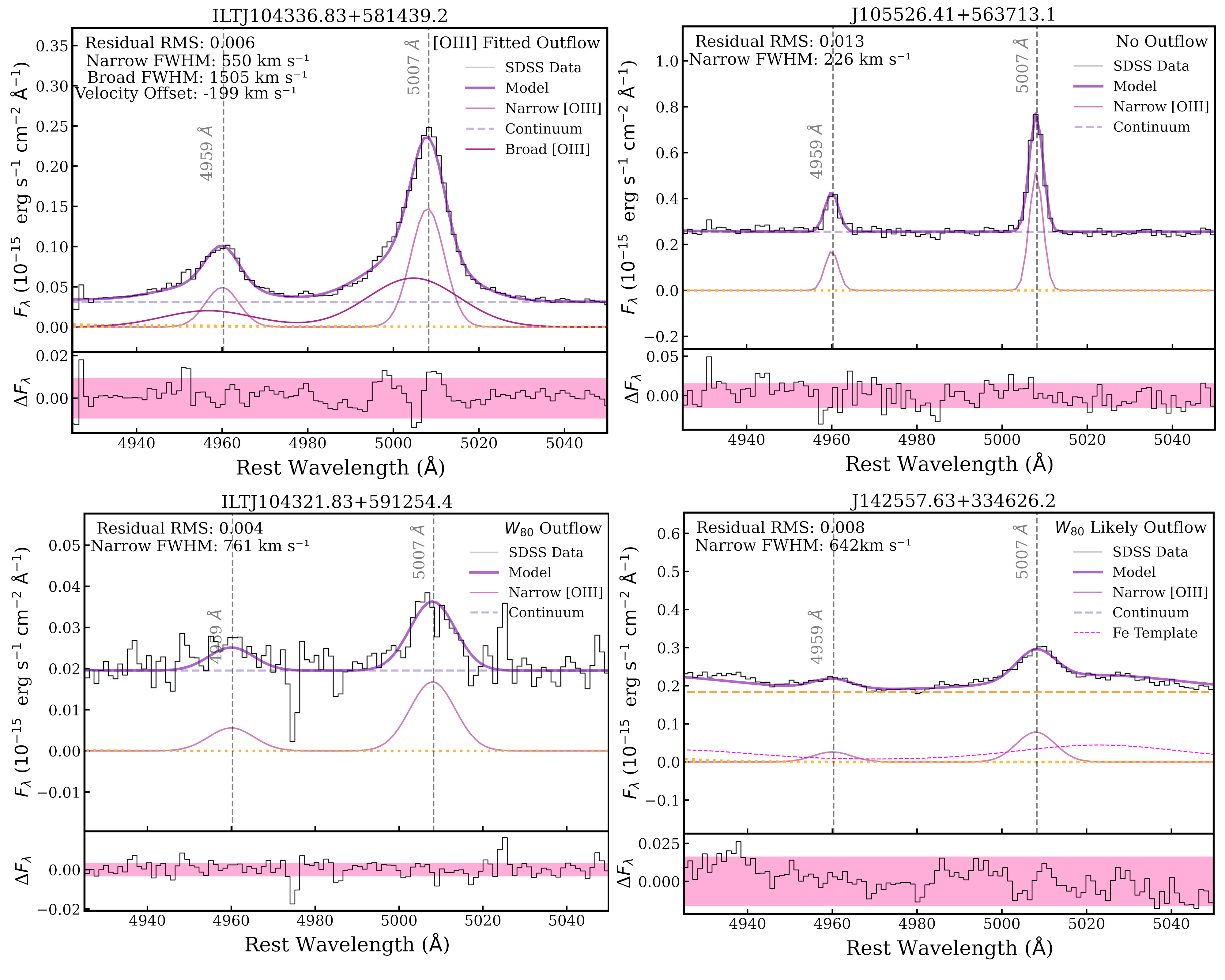}
    \caption{Example spectra for the four categories of the [O~{\sc iii}] SDSS spectral fitting results. The top panel of each subplot represents the SDSS spectral data in black, alongside the MCMC fitting results. The dark purple solid lines shows the final MCMC fitting result. The pink Gaussians represent the narrow component to [O~{\sc iii}] and the Gaussians in magenta, if plotted, is the broad component of [O~{\sc iii}] which implies an outflow is occurring. The light pink dashed lines, where present, shows the Fe II template used when modelling the data and finally the grey or yellow dashed lines shows the continuum. The lower panel of each subplot shows the residuals of the fit of the model to the SDSS data. The pink shaded region corresponds to rms 1$\sigmaup$ region, calculated over the full spectral range of the model. The titles of the radio detected AGN are the LoTSS source name of these AGN and start with ILTJ, the titles for the radio non-detected AGN are the SDSS name and they begin with J. Starting top left and going clockwise, the outflow groups of these AGN are [O~{\sc iii}] fitted outflow, no outflow, $W_{80}$ outflow, and $W_{80}$ likely outflow.}
    \label{fig:4_spectra}
\end{figure*}

After visual inspection of the spectral fits which are deemed to have the best fit by their BIC values, we find that for several spectra, the model with the lowest BIC value did not have the best fit to [O~{\sc iii}]. This is because we calculate the BIC value by considering the whole spectral window, i.e including H$\beta$. We therefore visually inspect all spectra to ensure that we prioritise the fit to the [O~{\sc iii}] emission line, rather than other emission lines such as H$\beta$ that are within the spectral window. Figure \ref{fig:4_spectra} shows some example spectra. \par

To calculate the [O~{\sc iii}] luminosity, we integrate the continuum-subtracted region between 4975 \AA\ and 5030 \AA\ . We then convert this to luminosity by using the distance modulus, using spectroscopic redshifts from SDSS. To calculate the error in the [O~{\sc iii}] luminosity we take the inverse square root of the inverse variance from the SDSS spectral information. We then sum the errors in quadrature. \par

\subsection{Identifying outflows} \label{identifing}

We identify an outflow if there is a second, blueshifted Gaussian fitted to [O~{\sc iii}] $\lambdaup$5007. However, even if an outflow is present, it might not always appear as a second component, due to observational limitations. Therefore, we further make use of the non-parametric $W_{80}$ measurement of [O~{\sc iii}] to identify outflows. $W_{80}$ is the velocity width containing 80 percent of the flux, calculated using the $\mathrm{10^{th}}$ and $\mathrm{90^{th}}$ percentiles of the velocities (${v_{10}}$ and ${v_{90}}$ respectively) of [O~{\sc iii}] $\lambdaup$5007 with ${W_{80}}$~=~${v_{90}}$~-~${v_{10}}$. We define two sub-categories of outflows following the definitions stipulated in \cite{harrison_kiloparsec-scale_2014}. When one component is fitted to [O~{\sc iii}], AGN with $W_{80}$~>~800~km~$\mathrm{s^{-1}}$ definitely host an outflow, and AGN with 600~km~$\mathrm{s^{-1}}$~<~$W_{80}$~<~800~km~$\mathrm{s^{-1}}$ are likely to be undergoing an outflow because the majority of the [O~{\sc iii}] total flux would be due to the outflowing component which is common especially in high luminosity AGN \citep[e.g.,][]{carniani_ionised_2015, kakkad_super_2020}. Applying these limits is necessary to ensure that the kinematics seen are due to outflowing gas rather than gas bounded by the NLR or produced by star formation. Therefore, we separate our sample into two categories of [O~{\sc iii}] outflows with sub-categories within them:

\begin{itemize}[align=parleft, left=\leftmargin]
    \item Outflow Present
    \begin{itemize}[labelwidth=\leftmargin, leftmargin=2\leftmargin]
        \item {[O~{\sc iii}]} Fitted Outflow - Second, blueshifted Gaussian present\\
        \item $W_{80}$ Outflow - One Gaussian fitted with $W_{80}$~>~800~km~$\mathrm{s^{-1}}$\\
        \item $W_{80}$ Likely Outflow - One Gaussian fitted with 600~km~$\mathrm{s^{-1}}$~<~$W_{80}$~<~800~km~$\mathrm{s^{-1}}$
    \end{itemize}
    \item No Outflow - One Gaussian fitted with $W_{80}$~<~600~km~$\mathrm{s^{-1}}$
\end{itemize}

An example of spectra showing each of these categories can be seen in Figure \ref{fig:4_spectra}, where the top left shows an [O~{\sc iii}] fitted outflow, and the bottom left panel presents a $W_{80}$ outflow, bottom right a $W_{80}$ likely outflow, and top right an AGN with no outflow. \par

\subsection{Stacking} \label{Stacking}

To help visualise the average proportions of the [O~{\sc iii}] emission line profile, we stack the radio detected population and radio non-detected populations separately within our matched sub-sample. As we are most interested in comparing the outflow properties of radio detected AGN to radio non-detected AGN, we first normalise each individual spectra by both their continuum as well as the peak of the narrow component of [O~{\sc iii}] and then stack these normalised spectra. By normalising the narrow component, we can directly compare the relative broad component in each stack. We use the stacking code presented in \cite{arnaudova_exploring_2024}. In the observed frame, we first correct the spectra for Galactic extinction by using the re-calibrated reddening data from \cite{schlegel_maps_1998}, along with the Milky Way reddening curve from \cite{fitzpatrick_correcting_1999}. Then, all spectra are shifted to the rest-frame by using the spectroscopic redshifts as measured from [O~{\sc iii}] and are resampled onto a common wavelength grid, with a channel width of 1 \AA\, using the \texttt{SpectRes}: Simple Spectral Resampling tool \citep{carnall_spectres_2017}. In order to perform the stacking procedure, we take the median of each of the spectra and normalise them by subtracting the median and then dividing by the peak of [O~{\sc iii}], where the median is computed at the reddest possible end of the area where all spectra populate the grid. A median stack is then performed where the associated uncertainties are bootstrapped. To ensure that the uncertainties are not underestimated, and that spectral features have not impacted the normalisation, an additional simulation is performed (see \cite{arnaudova_exploring_2024} for details).

\section{Results} \label{results} 

\subsection{Outflow Detection Rates} \label{detection rate}

\begin{figure}
    \centering
    \includegraphics[width=0.5\textwidth]{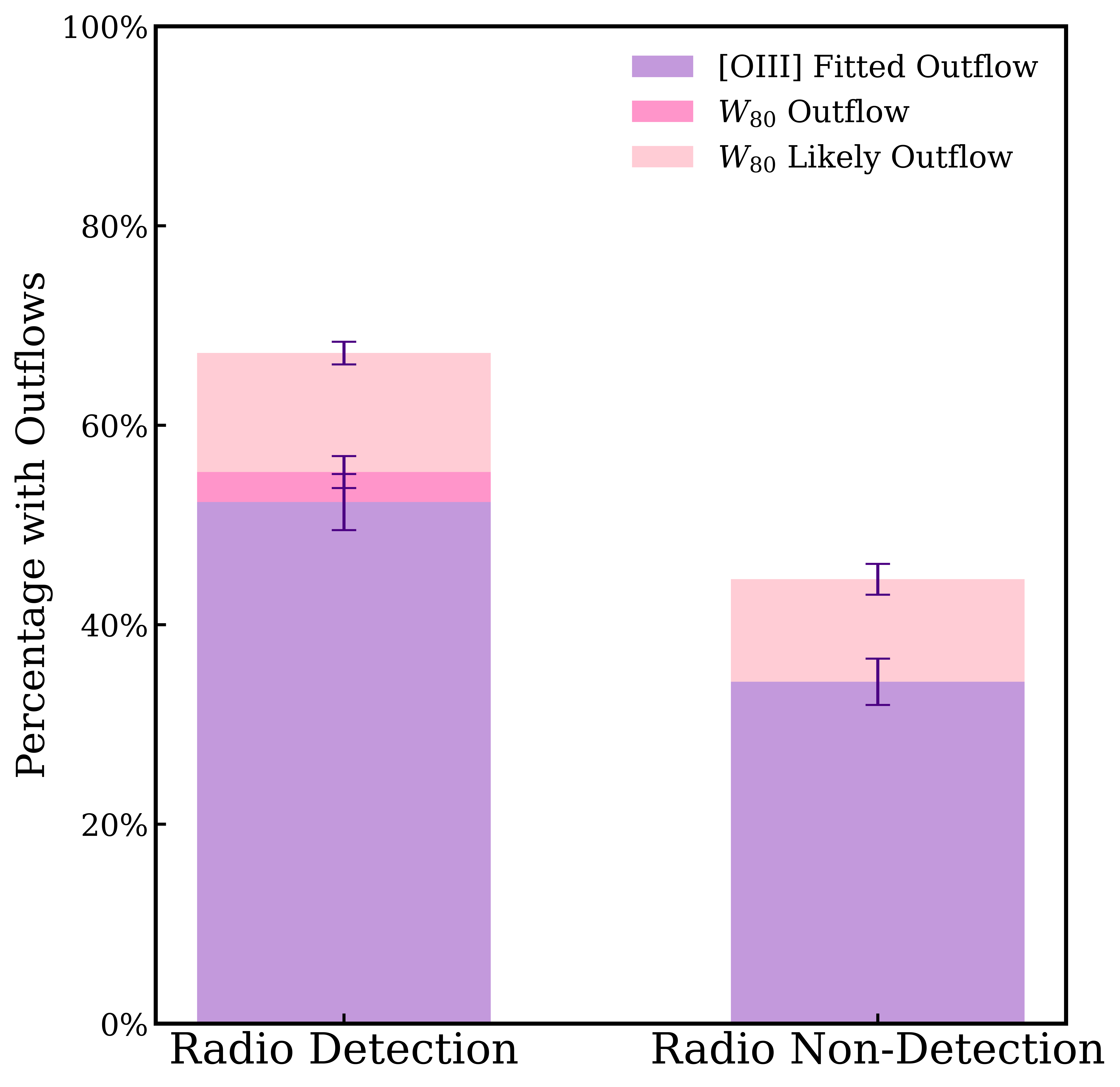}
    \caption{The stacked bar chart represents the average outflow detection rates from all 1000 iterations of randomly matching the radio detected AGN to the radio non-detected AGN. The light pink bar showing the $W_{80}$ likely outflows, the darker pink is the $W_{80}$ outflows, and the purple is the [O~{\sc iii}] fitted outflows. For the radio detected AGN, 52.3~$\pm$~2.8 percent have an [O~{\sc iii}] fitted outflow, 3.0~$\pm$~1.6 percent have $W_{80}$ outflows, 11.9~$\pm$~1.1 percent $W_{80}$ likely outflow, and 32.8~$\pm$~2.5 percent showing no signs of an outflow. In the same order the results for the radio non-detected AGN is, 34.3~$\pm$~2.3 percent, 0.0~$\pm$~0.0 percent, 10.3~$\pm$~1.5 percent and 55.4~$\pm$~2.4 percent.}
    \label{fig:precentage}
\end{figure}

\begin{figure}
    \centering
    \includegraphics[width=0.5\textwidth]{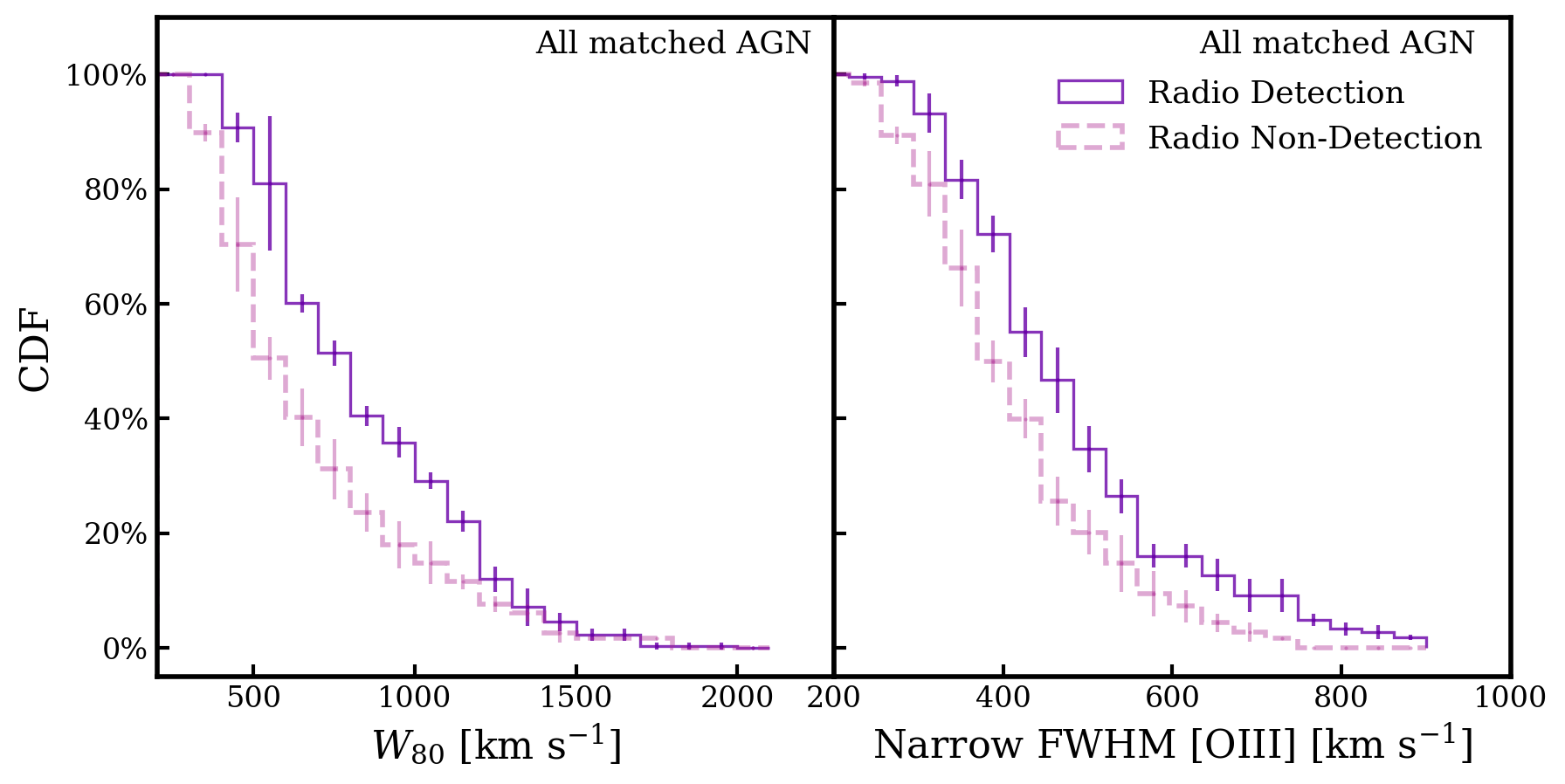}
    \caption{Average cumulative distribution functions of [O~{\sc iii}] properties. The solid purple step function shows the distribution for the radio detected AGN and the dashed pink line shows the information from the radio non-detected sources. $\emph{Left:}$ Average CDFs of $W_{80}$ [O~{\sc iii}] showing the results from all 1000 matching iterations. $\emph{Right:}$ Average CDFs of the FWHM of the narrow component of [O~{\sc iii}] with the all matching iterations. Uncertainties are constructed using bootstrapping.}
    \label{fig:two_culul}
\end{figure}

\begin{figure*}
    \centering
    \includegraphics[width=\textwidth]{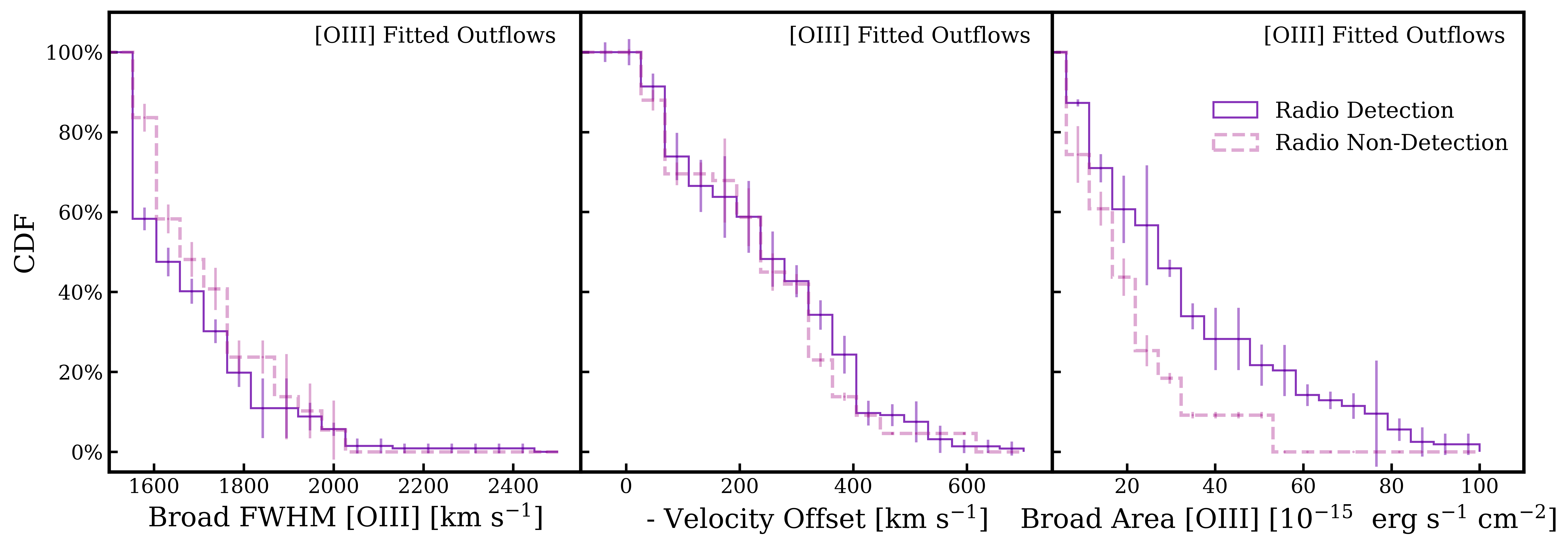}
    \caption{Further average cumulative distribution functions of [O~{\sc iii}] properties. The solid purple step function shows the distribution for the radio detected AGN and the dashed pink line shows the information from the radio non-detected sources. All three panels have AGN with a second component fitted to [O~{\sc iii}]. $\emph{Left:}$ CDF of the broad component of [O~{\sc iii}] FWHM. $\emph{Middle:}$ Negative velocity offset of the broad component relative to the narrow component. $\emph{Right:}$ Area of the broad component, calculated using the peak and FWHM of the broad component. Uncertainties are constructed using bootstrapping.}
    \label{fig:three_culul}
\end{figure*}

\begin{figure*}
    \centering
    \includegraphics[width=\textwidth]{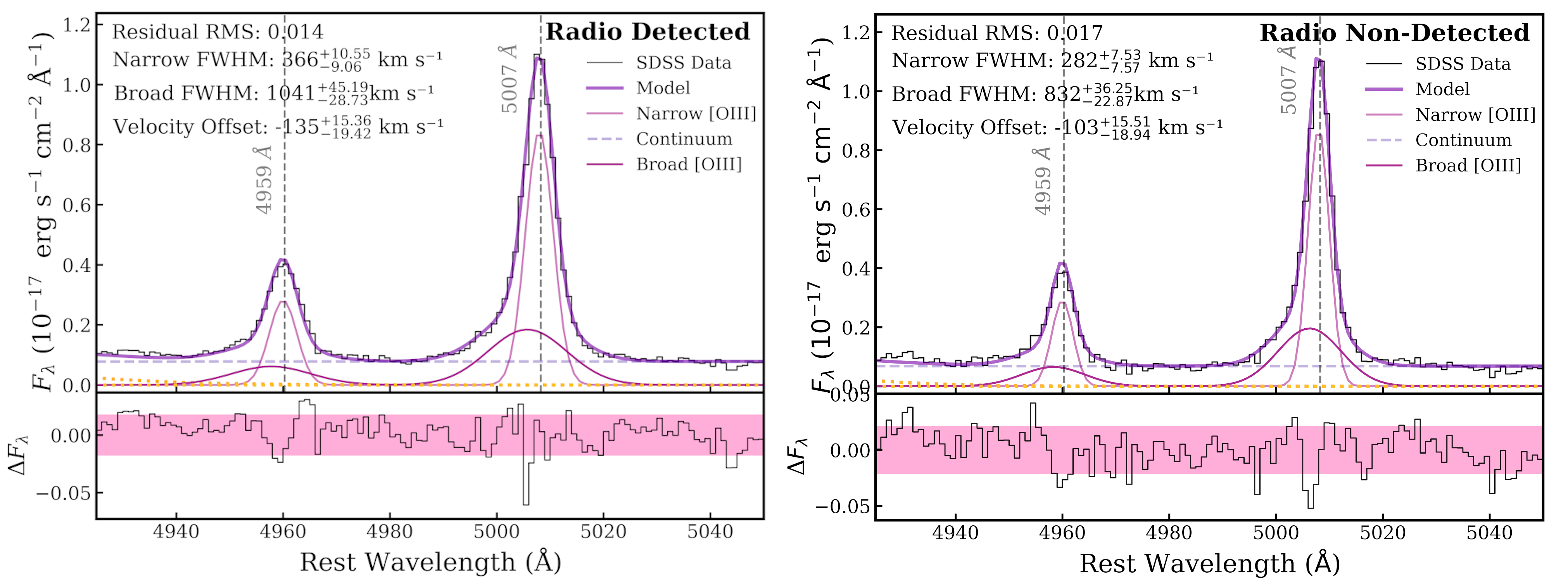}
    \caption{Composite spectra which are normalised by the peak of the narrow [O~{\sc iii}] component and are produced from the closest matched $L_{\mathrm{6\muup\\ m}}$ and $z$ run, containing 59 AGN in each population with the MCMC fitting tool results applied. $\emph{Left:}$ Radio detected AGN. $\emph{Right:}$ Radio non-detected AGN. The bottom panel shows the residuals of the fits with 1$\sigmaup$ of the residuals highlighted in the pink region. In the upper panel, the purple solid line shows the fit from the MCMC code to the SDSS data which is shown in back. The magenta line shows the broad component of [O~{\sc iii}] with the pink Gaussians representing the narrow component of [O~{\sc iii}].}
    \label{fig:stack}
\end{figure*}

\begin{figure*}
    \centering
    \includegraphics[width=\textwidth]{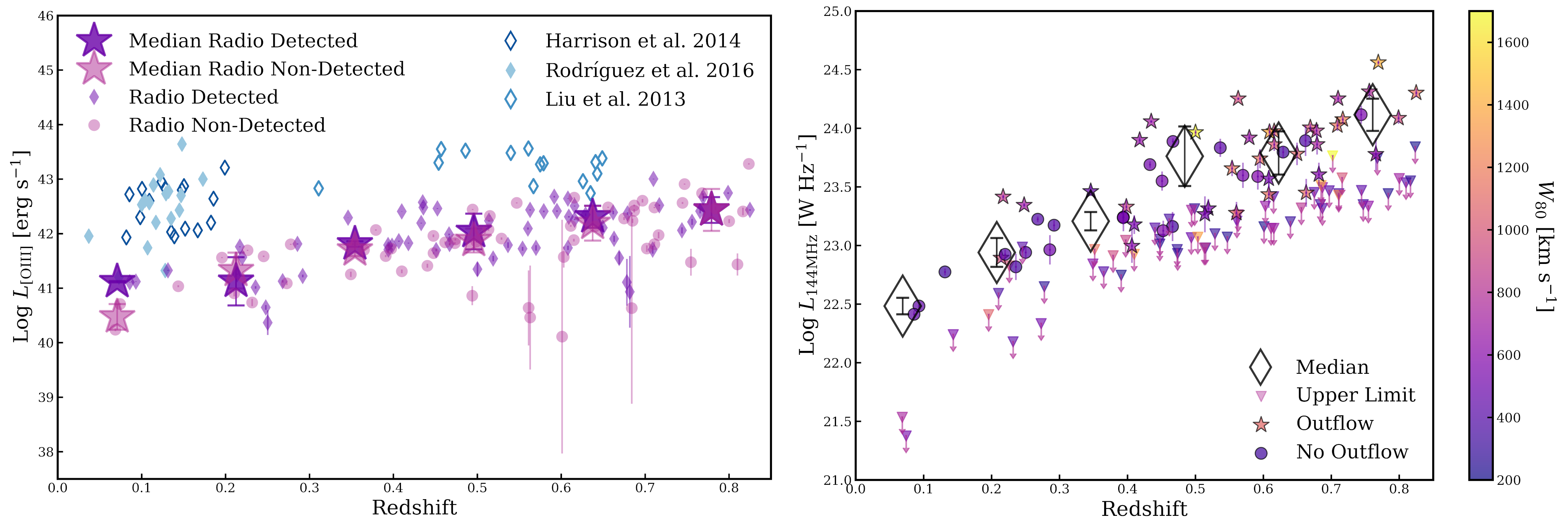}
    \caption{$\emph{Left:}$ Total $\lambdaup$5007 \AA\ ${L_{\mathrm{[OIII]}}}$ as a function of redshift. The star markers show the median values for six redshift bins, with purple indicating the median for the radio detected AGN, and pink for the radio non-detected AGN. The errors are the median absolute deviation. For the background sources, the purple diamonds show the radio detected population and the pink circles show the radio non-detected AGN. Similar studies are presented with various blue diamond markers. $\emph{Right:}$ Relationship between $L_{\mathrm{144MHz}}$ and redshift with the $W_{80}$ of [O~{\sc iii}] traced with a color map. Radio detected AGN are shown with either an outflow (small stars) or AGN with no outflow (circles). The median for five redshift bins are shown by large black hollow diamonds, with errors as the median absolute deviation. The upper limits for the radio non-detected AGN are shown by downward triangles.}
    \label{fig:lum_z}
\end{figure*}

\begin{figure}
    \centering
    \includegraphics[width=0.5\textwidth]{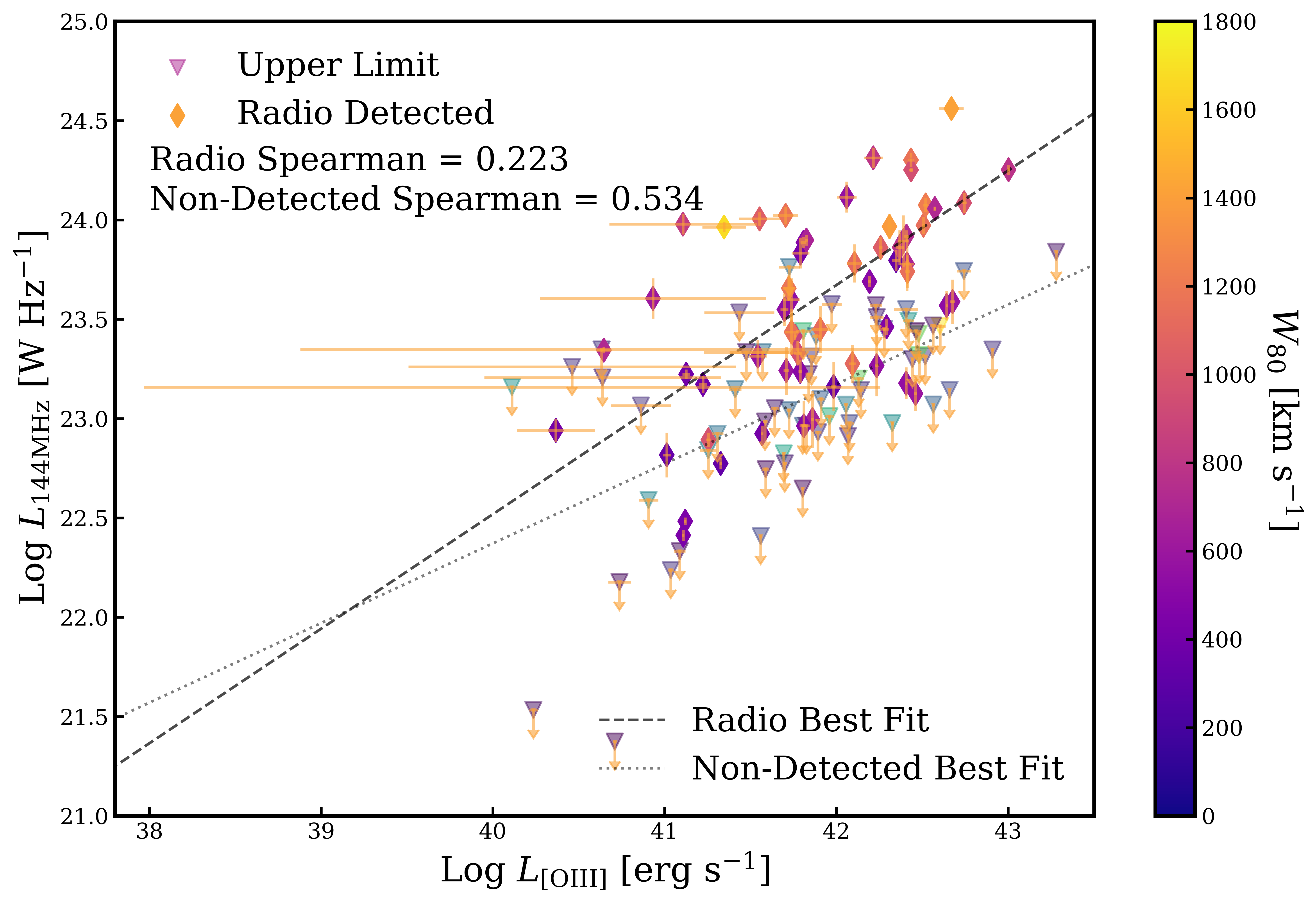}
    \caption{$L_{\mathrm{144MHz}}$ as a function of total ${L_{\mathrm{[OIII]}}}$ with the $W_{80}$ traced by a colour scale. Radio detected AGN are plotted with diamonds and the upper limits for the radio non-detected AGN are shown by the downward triangles. The dashed black line show the best fit relationship for the radio detected AGN and the grey dotted line shows the best fit to the upper limits of the radio non-detected population.}
    \label{fig:OIII_vs_radio}
\end{figure}

\begin{figure*}
    \centering
    \includegraphics[width=\textwidth]{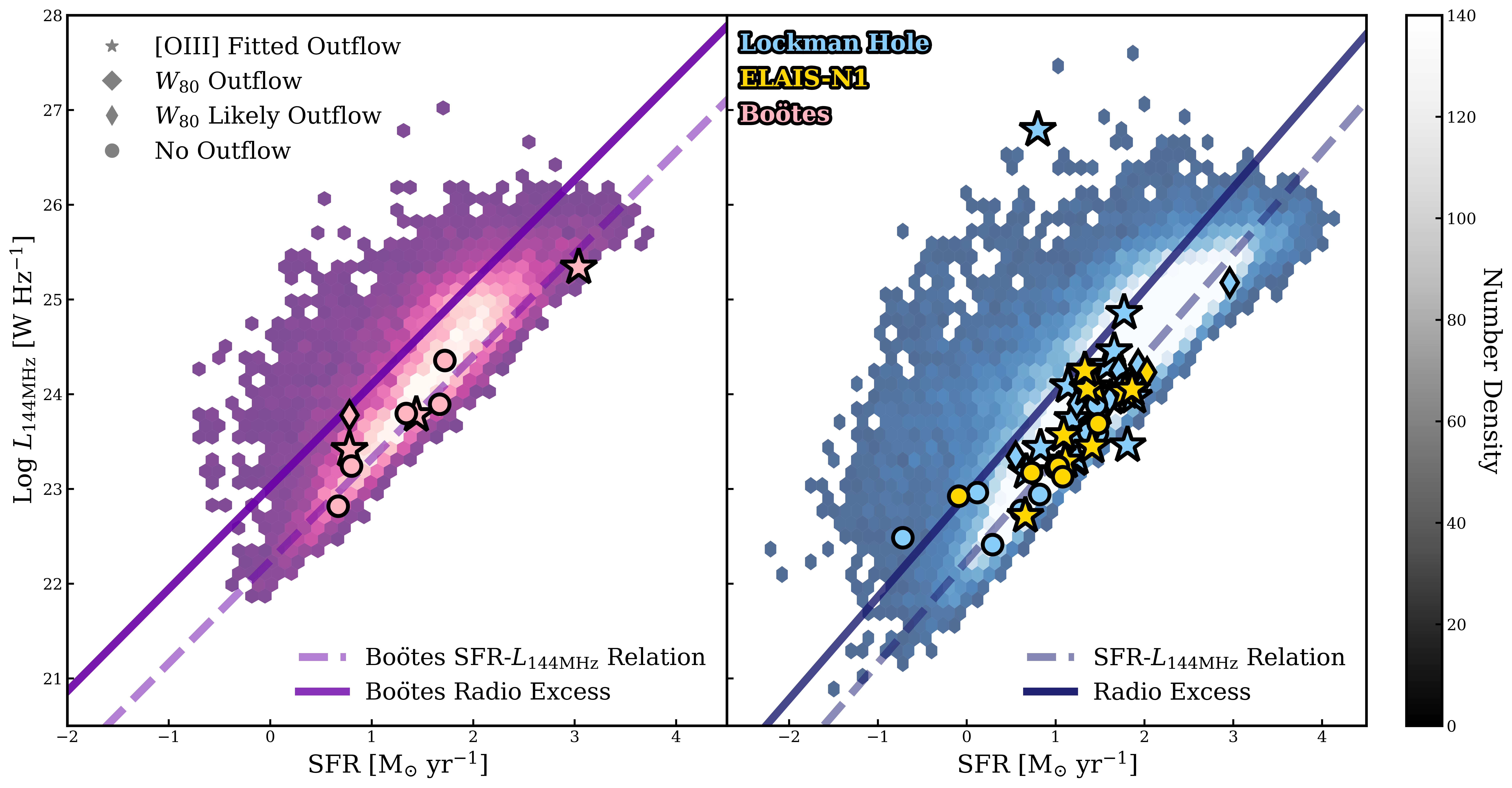}
    \caption{The relationship between SFR and $L_{\mathrm{144MHz}}$. The left subplot shows the result of the Bo\"{o}tes field (pink markers) and on the right Lockman Hole (blue markers) and ELAIS-N1 (yellow markers), the fields have been split because of the need for an extra adjustment to the radio excess in Bo\"{o}tes. Error bars are included but are too minimal to see. No upper limits are shown for the radio non-detected AGN because we do not have access for reliable SFR measurements for these AGN. The background hexagon bins show the distribution of sources with reliable SFR measurements with Bo\"{o}tes in purple and Lockman Hole and ELAIS-N1 combined in blue, where a hexagon is plotted for areas where there is four or more sources. On each panel, the faint dashed line shows the SFR-$L_{\mathrm{144MHz}}$ relation, which is the same for both fields. The solid line shows the radio excess divide which is 0.7 dex above the SFR-$L_{\mathrm{144MHz}}$ for Lockman Hole and ELAI-N1 with a slight redshift adjustment for Bo\"{o}tes. The over laid points with black edges are the sources within the best matched populations. The stars are sources with an [O~{\sc iii}] fitted outflow, thick diamond host an $W_{80}$ outflow and thin diamonds are AGN with a likely outflow. Circles represent sources that show no indication of an outflow. The colour scale traces the number density of each hexagon and is shown as grey as the scale is the same for both subplots, hence it would be pink for Bo\"{o}tes and blue for Lockman Hole and ELAIS-N1.}
    \label{fig:SFR}
\end{figure*}

\begin{figure}
    \centering
    \includegraphics[width=0.5\textwidth]{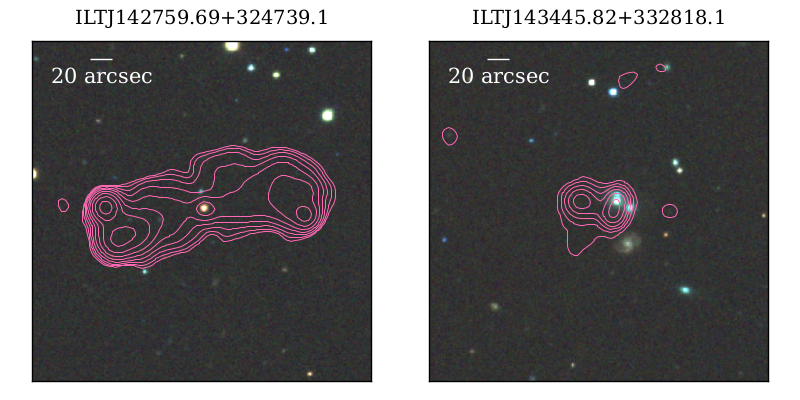}
    \caption{Cut outs (150\sarc\ x 150\sarc\ ) of the only two resolved Bo\"{o}tes radio sources within our sample. We use the $g,r$ and $i$ bands from SDSS to make a composite rgb optical image. The LoTSS radio contour plots are overlayed and the noise is set to 30$\muup$Jy \citep{tasse_lofar_2021}. The contour levels are 3$\sigmaup$, 5$\sigmaup$, 10$\sigmaup$, 20$\sigmaup$ and 40$\sigmaup$.}
    \label{fig:resolved}
\end{figure}

To obtain the [O~{\sc iii}] outflow detection rate, which is the fraction of AGN that host an outflow out of each population, we calculate the number of AGN which showed signs of each category of outflow, [O~{\sc iii}] fitted outflow, $W_{80}$ outflow, and $W_{80}$ likely outflow. As the matching process has inherent randomness, we calculate these fractions and bootstrap the uncertainties at the same time by re-performing the random $L_{\mathrm{6\muup\\ m}}$-$z$ matching 1000 times. This provides a more representative sampling of the data and its uncertainties. From the summation of these samplings, we calculate the mean value and its 1$\sigmaup$ (16th and 84th percentiles) uncertainties for each of these fractions. These errors are displayed in Figure \ref{fig:precentage}. \par

\begin{table}
    \centering
    \begin{tabular}{ccc}
    \hline
         & Radio Detected  & radio non-detected\\
         \hline
         \hline
        $[$OIII$]$ Fitted Outflow & 52.3 $\pm$ 2.8  & 34.3 $\pm$ 2.3\\
        $W_{80}$ Outflow & 3.0 $\pm$ 1.6 & 0.0 $\pm$ 0.0\\
        $W_{80}$ Likely Outflow & 11.9 $\pm$ 1.1  & 10.3 $\pm$ 1.5\\
        All Outflow Categories & 67.2 $\pm$ 3.4  & 44.6 $\pm$ 2.7 \\
        Definite Outflows & 55.3 $\pm$ 3.2 & 34.3 $\pm$ 2.3 \\
        No Outflow & 32.8 $\pm$ 2.5 & 55.4 $\pm$ 2.4 \\
        \hline
    \end{tabular}
    \caption{Outflow detection rates of the radio detected and radio non-detected populations, where [O~{\sc iii}] fitted outflows have two components fitted to [O~{\sc iii}], $W_{80}$ outflows have a single component and $W_{80}$~>~800~km~$\mathrm{s^{-1}}$, $W_{80}$ likely outflows have a single component and 600~km~$\mathrm{s^{-1}}$~<~$W_{80}$~<~800~km~$\mathrm{s^{-1}}$, and no outflows where a single component is fitted to [O~{\sc iii}] with $W_{80}$~<~600~km~$\mathrm{s^{-1}}$. AGN within the Definite outflow category have an outflow classed as either a [OIII] fitted outflow, or a $W_{80}$ outflow. AGN classed with a [O~{\sc iii}] fitted outflow, $W_{80}$ outflow or $W_{80}$ likely outflow are consider in the all outflow category.}
    \label{tab:outflow_rate}
\end{table}

Figure \ref{fig:precentage} clearly shows that we see a higher outflow detection rate in radio detected AGN than radio non-detected AGN, with >3$\sigmaup$ significance for AGN with a fitted [O~{\sc iii}] outflow. Considering all outflow categories, 67.2~$\pm$~3.4 percent of radio detected AGN indicate signs of an outflow compared to just 44.6~$\pm$~2.7 percent of the radio non-detected AGN. If we consider only definite outflows within the [O~{\sc iii}] fitted and $W_{80}$~>~800~km~$\mathrm{s^{-1}}$ categories, the outflow detection rates are 55.3~$\pm$~3.2 percent for the radio detected AGN but 34.3~$\pm$~2.3 percent for the radio non-detected AGN. \par

\subsection{[O~{\sc iii}] Properties} \label{cumul}

To further understand the differences between [O~{\sc iii}] from radio detected AGN and radio non-detected AGN, we plot the cumulative distribution functions (CDFs) of average emission line properties in Figures \ref{fig:two_culul} and \ref{fig:three_culul}. As for the outflow detection rates, we use the 1000 matching runs to bootstrap the uncertainties. These CDFs encompass all 1000 matching runs. For the uncertainties, we bootstrap over all these 1000 matching runs to obtain a distribution of AGN for each bin. From these distributions we fit a Gaussian and take the standard deviation as the presented uncertainty of each bin.\par

Figure \ref{fig:two_culul} shows how the populations differ in $W_{80}$ and the FWHM of the narrow, or single Gaussian component fitted to [O~{\sc iii}]. These CDFs contain all matched AGN. The CDF of $W_{80}$ (Figure \ref{fig:two_culul} left) shows us that the radio detected population has a significantly larger $W_{80}$ across the majority of the population than in the radio non-detected sample. The results change slightly when considering the AGN within the [O~{\sc iii}] fitted outflow category, where radio detected AGN show a significantly larger $W_{80}$ at intermediate bins (See Appendix A for additional CDFs). We do point out however that $W_{80}$ is a non-parametric measurement and is degenerate between the narrow and the broad components. \par

We check if the $W_{80}$ is dominated by one component by examining if there is a clear relationship between $W_{80}$ and either the broad or narrow components. There is a clear dependency only when a single component is present, as expected. This is consistent with $W_{80}$ being a non-parametric measurement with degeneracies between two components when they are present. \par

The right panel in Figure \ref{fig:two_culul} presents the CDFs of the FWHM of the narrow, or single Gaussian component of the [O~{\sc iii}] $\lambdaup$5007 emission line. It appears that radio detected AGN significantly dominate in the vast majority of bins when all matched AGN are considered. Radio detected AGN with two components fitted to [O~{\sc iii}] still seem to dominate however at a lesser extent. For AGN with a single component fitted to [O~{\sc iii}] the radio detected AGN appear to be very dominant within this category at all levels of FWHM (see Appendix A). \par

The left panel of Figure \ref{fig:three_culul} presents the CDF of the FWHM of the broad component which shows that radio non-detected AGN have a larger FWHM of the second component compared to the radio detected AGN until the highest FWHM where the distributions become equal within the uncertainties. This implies that if a second component is fitted to [O~{\sc iii}], then this would be broader if the AGN does not have a detection in LoTSS. \par

The velocity offset CDF (middle panel in Figure \ref{fig:three_culul}) shows us that the two populations have similar velocity offsets or blueshifts, aside from a few intermediate bins, where the radio detected AGN significantly dominate. We can therefore imagine that the two components in radio detected AGN would have similar separation as the two components in radio non-detected AGN. \par

Finally the right panel in Figure \ref{fig:three_culul} shows the CDF for the area of the broad component of [O~{\sc iii}] for both the radio detected and radio non-detected populations. This area is calculated using both the peak and FWHM of the broad component. This CDF informs us that radio detected AGN have a brighter broad, second, component than radio non-detected AGN at a significant level for the majority of the bins. \par

One of the main differences we see in the CDFs between the two populations is the kinematics of the narrow, [O~{\sc iii}] core component with the radio detected AGN having a larger FWHM. As we are most interested in the difference between the [O~{\sc iii}] outflowing properties of the two populations, we produce a stack in Figure \ref{fig:stack} where we show the spectral fitting conducted to the median stacked spectra after we normalised spectra in the stack by the peak of the narrow component. The AGN contributing to these figures are from the closest matched population of the $L_{\mathrm{6\muup\\ m}}$ and redshift matching process. The left panel shows the results of the radio detected AGN and on the right we see the radio non-detected AGN. We see that both the radio and radio non-detected AGN stacking results show that there appears to be an [O~{\sc iii}] fitted outflow in both populations as a broad, second, component is fitted to both. It is interesting to note that from Figure \ref{fig:stack} we see an enhanced outflowing component in the normalised stack for the radio detected AGN. To quantify this, we calculate the area of just the broad component and find that the radio detected AGN have an outflowing component with an area of 204.74~$^{{+{17.97}}}_{{-{18.98}}}$~$~\mathrm{km^{2}}~$$\mathrm{s^{-2}}$ compared to the radio non-detected AGN having an outflowing area of 173.91~$^{{+{17.74}}}_{{-{16.11}}}$~$~\mathrm{km^{2}}~$$\mathrm{s^{-2}}$. We note that this result is not significant when considering the extremities of these uncertainties by <5~$\mathrm{km^{2}}~$$\mathrm{s^{-2}}$. As radio detected AGN show a brighter second component after normalisation of the narrow component, this indicates that this is not a luminosity dependent affect because the ratio of the broad to the narrow component is larger for radio detected AGN than radio non-detected AGN, regardless of the brightness of the narrow component. From these stacks, we find the velocity offset is larger for the radio detected AGN, however if we consider the extremity of the uncertainties, this difference is not significant by <3~km~$\mathrm{s^{-1}}$. \par

Combining all the information from the CDFs and the stacked spectra, we can begin to build a picture of [O~{\sc iii}] for typical radio and radio non-detected AGN. The [O~{\sc iii}] $\lambdaup$5007 of radio detected AGN would appear to have a slightly broader, narrow component, but a narrower broad component than radio non-detected AGN, with a brighter broad component of [O~{\sc iii}]. The distance between these two components would be similar for both populations. \par

\subsection{Radio and [O~{\sc iii}] Luminosity Relationship} \label{radio OIII relation}

In this section, we explore the relationship between radio and [O~{\sc iii}] luminosity shown in Figure \ref{fig:lum_z}. The left panel shows the [O~{\sc iii}] $\lambdaup$5007\ luminosity as a function of redshift. We bin our results in six redshift bins and take the median value of $z$ and ${L_{\mathrm{[OIII]}}}$ for the AGN in each bin. Other work conducted with similar redshift ranges are also shown \citep{harrison_kiloparsec-scale_2014, liu_observations_2013, rodriguez_zaurin_importance_2013}. Our sample has lower [O~{\sc iii}] luminosities on average than these previous studies. \par

The right panel in Figure \ref{fig:lum_z} depicts the relationship between $L_{\mathrm{144MHz}}$ and redshift with the $W_{80}$ of [O~{\sc iii}] traced with a colour scale, with $L_{\mathrm{144MHz}}$ of the radio non-detected AGN being upper limits. We create five redshift bins and the median values of $L_{\mathrm{144MHz}}$ for each bin are shown, with the median absolute deviation as the error. \par

The relationship between [O~{\sc iii}] and radio luminosity was first presented in \cite{rawlings_relations_1989}. They discovered a positive correlation between the total ${L_{\mathrm{[OIII]}}}$ (both 5007 \AA\ and 4959 \AA) and $L_{\mathrm{178MHz}}$. Figure \ref{fig:OIII_vs_radio} shows $L_{\mathrm{144MHz}}$ as a function of ${L_{\mathrm{[OIII]}}}$ (5007 \AA\ only) with $W_{80}$ traced with a colour scale. Here we see that as $L_{\mathrm{144MHz}}$ increases, so does ${L_{\mathrm{[OIII]}}}$, which is to be expected and is in agreement with \cite{rawlings_relations_1989}. We note that our results show a Spearman rank correlation of 0.223, whereas \cite{rawlings_relations_1989} presents a value of 0.51. A value of zero indicates no association, while a value of unity indicated a strong association. We therefore find a weaker correlation than \cite{rawlings_relations_1989}. The difference may be due to \cite{rawlings_relations_1989} using the total ${L_{\mathrm{[OIII]}}}$ while we use just $\lambdaup$5007 \AA\ to calculate ${L_{\mathrm{[OIII]}}}$. Our results produce a best fit with a slope of 0.58. \cite{rawlings_relations_1989} does not remove any relation that would be introduced by an AGN luminosity bias, however even after matching in $L_{\mathrm{144MHz}}$ and redshift, we still see a correlation between [O~{\sc iii}] and radio luminosity. We note that we can not determine if this relationship is real or due to the effect of the limited field of view and flux limits of our sample. To determine this we would require a larger sample.\par 

\subsection{Radio Emission} \label{radio_emission}
\subsubsection{Radio Excess} \label{radio_excess}

To investigate the origin of the radio emission, we show the relationship between SFR and $L_{\mathrm{144MHz}}$ in Figure \ref{fig:SFR}. Only radio detected sources are shown in this figure since \cite{best_lofar_2023} only carried out SED fitting to calculate the SFR for radio detected sources (see Section \ref{muliti-data}). We use the redshift presented in \cite{best_lofar_2023}, which can be either spectroscopic or high quality photometric redshifts from \cite{duncan_lofar_2021}, to calculate the radio luminosities to be consistent with their results. Overlaid we show our $L_{\mathrm{6\muup\\ m}}$-$z$ matched sample.\par 

On the left we see the results for the sources within the Bo\"{o}tes field and on the right we see the Lockman Hole and ELAIS-N1 sources. We show these fields separately as the scatter in the Bo\"{o}tes is slightly different, leading to an extra adjustment in this field. We show the SFR and $L_{\mathrm{144MHz}}$ relation found in \cite{best_lofar_2023} in both panels as well as the radio excess line which is 0.7 dex above the relation in Lockman Hole and ELAIS-N1. For Bo\"{o}tes an extra redshift dependent correction is necessary so the radio excess line is 0.7+0.1$z$ above the relation, due to increased scatter in the SFR and $L_{\mathrm{144MHz}}$ relation for this field. Above these lines, sources are considered to be radio excess, where their radio emission is dominated by star formation. To guide the eye, we use the maximum redshift of our sample, $z$~=~0.83, for the redshift dependent definition of radio excess in the Bo\"{o}tes field. \par

Only four sources from our sample appear to be radio excess as they lie above the radio excess divide, one in ELAIS-N1, three in Lockman and none in Bo\"{o}tes. Two of the radio excess AGN have a broad, second component fitted to [O~{\sc iii}]. Radio emission in sources above the radio excess line can be attributed to radio jets. The majority of our sample lies below the radio excess line, and we are unable to use this diagnostic to determine the origin of their radio emission. \par

\subsubsection{Radio Morphology} \label{radio_morp}

Radio morphology can help diagnose the source of radio emission, for example by identifying jets which can only be produced by AGN and not star formation. We study the radio morphology of the radio detected population at 6\sarc\ which is available using the LoTSS catalogue. We visually classify the morphology of our sample. In all our 115 radio detected AGN, we find $\sim$93 percent of sources (107/115) are unresolved, meaning only eight sources show resolved structure; two of these resolved AGN which lie within Bo\"{o}tes are shown in Figure \ref{fig:resolved}, five in Lockman, and one in ELAIS-N1. These unresolved source imply radio sizes less than 4.7 kpc. \par

For the majority of our sample, we therefore cannot use morphology to determine the source of the radio emission. This emphasises the need for high resolution imaging to securely identify the radio morphology.\par

\subsection{Comparison to Other Studies} \label{compare}

The [O~{\sc iii}] emission line kinematics in radio enhanced vs. radio weak AGN were studied extensively by \cite{mullaney_narrow-line_2013}. The authors show that AGN with $L_{\mathrm{1.4GHz}}$~>~$\mathrm{10^{23}}$~W~$\mathrm{Hz^{-1}}$ have a larger average FWHM than AGN with lower radio luminosities. In this large statistical sample, the authors use the flux-weighted average FWHM of the broad and narrow components whereas we study them separately. Although we have a smaller sample size, our study extends to lower radio luminosities (down to $L_{\mathrm{1.4GHz}}\sim\mathrm{10^{20}}$~W~$\mathrm{Hz^{-1}}$) and higher redshifts, and we match in $L_{\mathrm{6\muup\\ m}}$ and redshift which may contribute to the seemingly different results we find. However, overall conclusions are still consistent: our sample is not dominated by powerful radio jets (Section \ref{radio_excess}), and we find differences in sources with strong vs. weak radio emission \citep[either radio detected vs. not radio detected, this work; or $L_{\mathrm{1.4GHz}}$~>~$10^{23}$ vs. $L_{\mathrm{1.4GHz}}$~>~$10^{23}$~W~$\mathrm{Hz^{-1}}$,][]{mullaney_narrow-line_2013}. To check the consistency of our samples we convert our $L_{\mathrm{144MHz}}$ to $L_{\mathrm{1.4GHz}}$ assuming a spectral index of 0.7 and split the radio detected sample into sources with $L_{\mathrm{1.4GHz}}$~>~$\mathrm{10^{23}}$~W~$\mathrm{Hz^{-1}}$ or $L_{\mathrm{1.4GHz}}$~<~$\mathrm{10^{23}}$~W~$\mathrm{Hz^{-1}}$ as well as removing sources that are above a redshift of 0.4 to match their selection. We then calculate the average FWHM as in Equation 2 from \cite{mullaney_narrow-line_2013}, and find the same trend for the average FWHM for low and high luminosity AGN \cite[see Fig. 9 in][]{mullaney_narrow-line_2013}. \par

Recently \cite{calistro_rivera_multiwavelength_2021} and \cite{calistro_rivera_ubiquitous_2023} found that red quasars \cite[which are known to be radio enhanced;][]{klindt_fundamental_2019} have a larger FWHM of the second component fitted to [O~{\sc iii}] than their control quasars. It is difficult to draw a direct comparison between their work and ours because our sample spans the whole range of colours and the majority would not be classed as red quasars. \par

\section{Discussion} \label{discussion}

We see that radio detected AGN have a higher outflow detection rate compared to our radio non-detected population as well as differences in the kinematics of the [O~{\sc iii}] emission line between the two populations. After normalising for the narrow [O~{\sc iii}] component, in the resulting stacked spectra we see that the radio detected AGN have a larger broad outflowing component than the radio non-detected AGN. This is suggestive of the radio detected AGN hosting outflows with a larger amount of gas than outflows produced by the radio non-detected population.\par

\subsection{What is the Origin of Radio Emission?}\label{origin}

The SFR-$L_{\mathrm{144MHz}}$ relation has shown us that the majority of sources are not radio excess in both fields. Sources above the radio excess divide have radio emission which is likely to originate from a powerful jet. As only four AGN are above the radio excess divide, this shows that the radio emission is not driven by powerful radio jets. However we can only rule out powerful jets as the dominant cause of the radio emission in this sample. The results of this paper also indicate that radio emission is produced when a large amount of ionised outflowing gas is present and here we discuss two possible explanations as to why we see this. The possible mechanisms responsible for the radio emission in this sample are star formation, small (weak) radio jets, or shocks associated with AGN winds.\par

Star formation is unlikely to be the driver of radio emission in our sample. We see that in radio detected AGN, there is a significantly larger amount of outflowing gas. Star formation is a relatively weak and widespread phenomenon in a galaxy, and therefore is unlikely to be the main driving mechanism for the increase in outflow gas. \par 

Radio emission from low-power or small scale jets would be directional, and have a particular geometric relationship with ionised outflows. If the radio jets are co-spatial with the outflows, then it is possible the correlation between outflow detection rate and radio detection is due to orientation effects. In this scenario, the radio emission could be higher due to beaming in sources where outflows are along our line of sight. However, at the low-frequencies of these observations, we expect beaming effects to be negligible. Small scale jets can also explain the increased gas level as it could lead to more fuel for the AGN, and this fuel can possibly be launched as a radio jet later on in the AGN life cycle. \par

The alternative explanation is that AGN winds can be the main mechanism in producing radio emission in our sample. If the outflow is in the form of a wind then the higher level of gas would lead to the wind becoming more dense, and this increased density could cause a shock to occur within the wind and hence produce radio emission due to the particle acceleration within the shock front. This could explain the higher level of outflowing gas seen in radio detected AGN. \cite{zakamska_quasar_2014} suggests that the driving mechanism is quasar winds that propagate into the ISM and produce shock fonts where particle acceleration occurs, hence producing radio emission. However, the authors also acknowledge that the radio and [O~{\sc iii}] correlation could be driven by the mechanical energy of relativistic jets due to excessive heating releasing an ionised wind. The shocks suggested by \cite{zakamska_quasar_2014} are not expected to produce radio emission causing sources to be radio excess. \cite{nims_observational_2015} calculates that the synchrotron emission generated by shocks are about 5 percent of the bolometric luminosity and in a later study \cite{zakamska_discovery_2016} investigates the kinematics of [O~{\sc iii}] of four extremely red quasars and estimates that at least 3 percent of the bolometric luminosity is powering the kinematic energy of dusty winds. The addition of these to the radio emission from star formation is still well within the scatter of the SFR-$L_{\mathrm{144MHz}}$ relation, and the fact that a source is not radio excess does not rule out that AGN wind shocks are the cause of the radio emission. Therefore, high resolution follow-up will be crucial to determine the origin of this radio emission.

The driving mechanism linking radio emission and [O~{\sc iii}] outflows is still up for debate. As already discussed, \cite{zakamska_quasar_2014} suggests that utflows trigger shocks which drive the radio emission. More Recently, \cite{liao_outflow-related_2024} studies the connection between $w_{90}$ of [O~{\sc iii}] and radio emission from median stacked radio images from the Karl G. Jansky Very Large Array (VLA) Sky Survey (VLASS) \citep{lacy_karl_2020} from $\sim$37,000 radio quiet AGN. Similar to this work, the authors confirm a significant connection between [O~{\sc iii}] and radio emission as well as suggesting the most likely explanation for the radio emission from their radio quiet AGN is also AGN driven by either low-powered jets or winds. \par

\cite{jarvis_prevalence_2019,jarvis_quasar_2021} follows-up around 40 radio quiet AGN from \cite{mullaney_narrow-line_2013} with high frequency, high resolution radio images from the VLA and e-MERLIN. These high resolution images proved to be crucial to determining the origin of radio emission from this sample of radio quiet AGN, where the authors found that the majority of the radio emission originates from radio jets with star formation likely to be responsible for $\sim$10 percent of the radio emission. This highlights that high resolution imaging of our sample will be key to furthering our understanding of the radio emission from our sample of 198 AGN. \par

\section{Conclusions} \label{conclusion}

We form a sample of 198 AGN in the LoTSS Deep Fields, 115 of which are detected in LoTSS at 144MHz, and 83 with no detection. We use SDSS as a base sample and supplement the data with the deep multi-wavelength information from the LoTSS Deep Fields. We match the radio non-detected to the radio detected AGN in $L_{\mathrm{6\muup\\ m}}$ and redshift to remove any possible biases from AGN luminosity. \par 

Using an MCMC statistical approach, we fit the [O~{\sc iii}] emission line. We place each AGN in one of the four [O~{\sc iii}] outflow sub-categories: i) [O~{\sc iii}] fitted outflows, where two Gaussians are fitted to [O~{\sc iii}]; ii) $W_{80}$ outflow with one component fitted and $W_{80}$~>~800~km~$\mathrm{s^{-1}}$; iii) $W_{80}$ likely outflows with one component fitted and 600~km~$\mathrm{s^{-1}}$~<~$W_{80}$~<~800~km~$\mathrm{s^{-1}}$; iv) and finally no outflow, where one component is fitted to [O~{\sc iii}] and $W_{80}$ < 600~km~$\mathrm{s^{-1}}$. \par

In the radio detected population we find 67.2~$\pm$~3.4 percent of these AGN show signs of an outflow occurring compared to just 44.6~$\pm$~2.7 percent of radio non-detected AGN. Even when we remove the $W_{80}$ outflow based sub-categories, radio detected AGN still show a higher outflow detection rate. This indicates that [O~{\sc iii}] outflows are more common in AGN where there is a detection in LoTSS compared to there being no detection present. \par

We use both stacked spectra and CDFs of the emission line parameters to understand the average profile of [O~{\sc iii}] of the two populations. The CDFs tell us that radio detected AGN have a larger $W_{80}$ than the radio non-detected AGN as well as a broader, [O~{\sc iii}] core component, however radio non-detected AGN have a broader, second component. To allow us to just study the differences between the outflowing component [O~{\sc iii}], we produce stacks which have been normalised by the peak of the narrow component and find that radio detected AGN have a larger outflowing component with respects to the radio non-detected AGN, implying that more gas is present in [O~{\sc iii}] outflows hosted by a radio detected AGN. This suggests radio emission from either low-powered jets or shocks from AGN driven winds. \par 

We find that the majority of our radio detected sources are not radio excess. Therefore this radio emission appears not to originate from high-powered radio jets but is more likely from star formation, winds or low-powered radio jets. \par

To determine the origin of this radio emission from radio quiet AGN, we must study the radio morphology. However, we find that $\sim$93 percent of sources within all three LoTSS Deep Fields are unresolved at 6\sarc\ . By incorporating the international stations of LOFAR we can obtain sub-arcsecond resolutions down to 0.3\sarc\ which improves the resolution by a factor of 20. Using this sub-arcsecond resolution imaging, which provides access to sub-galactic scales, we can use radio morphology and brightness temperature to determine the nature of AGN emission and separate it from radio emission due to star formation \citep{morabito_sub-arcsecond_2022}. With this sub-arcsecond resolution we will be able to study the morphology of previously unresolved sources. This will allow us to determine if the radio emission from these AGN is from the AGN activity or star formation. \par 

The Lockman Hole Deep Field is the first widefield image produced at sub-arcsecond resolution with a frequency of $L_{\mathrm{144MHz}}$ \citep{sweijen_deep_2022}. This process is very computationally heavy and took an estimated 250,000 core hours. A deeper image of ELAIS-N1, at the same resolution and frequency, combines four nights of observations, has also been released in \cite{de_jong_into_2024}. Using the same techniques, we are processing the Bo\"{o}tes Deep Field which will result in another widefield image at sub-arcsecond resolution at $L_{\mathrm{144MHz}}$. With these three datasets, we will be able to study their radio morphology and determine whether the radio emission is from AGN activity or star formation. This will lead us closer to understanding the nature of radio emission in radio quiet AGN. \par

\section*{Acknowledgements}

EE and LKM are grateful for support from the Medical Research Council [MR/T042842/1], CMH acknowledges funding from UKRI (MR/V022830/1), MIA acknowledges support from the UK Science and Technology Facilities Council (STFC) studentship under the grant ST/V506709/1, RCH acknowledges support from NASA through ADAP grant number 80NSSC23K0485, RK acknowledges support from the STFC via grant ST/V000594/1 and DJBS acknowledges support from the UK STFC under grant ST/V000624/1. \par

This paper is based on data obtained with the International LOFAR Telescope (ILT). LOFAR \citep{haarlem_lofar_2013} is the Low Frequency Array designed and constructed by ASTRON. It has observing, data processing, and data storage facilities in several countries, that are owned by various parties (each with their own funding sources), and that are collectively operated by the ILT foundation under a joint scientific policy. The ILT resources have benefited from the following recent major funding sources: CNRS-INSU, Observatoire de Paris and Université d'Orléans, France; BMBF, MIWF-NRW, MPG, Germany; Science Foundation Ireland (SFI), Department of Business, Enterprise and Innovation (DBEI), Ireland; NWO, The Netherlands; The Science and Technology Facilities Council, UK; Ministry of Science and Higher Education, Poland; The Istituto Nazionale di Astrofisica (INAF), Italy. \par

This research made use of the Dutch national e-infrastructure with support of the SURF Cooperative (e-infra 180169) and the LOFAR e-infra group. The Jülich LOFAR Long Term Archive and the German LOFAR network are both coordinated and operated by the Jülich Supercomputing Centre (JSC), and computing resources on the supercomputer JUWELS at JSC were provided by the Gauss Centre for Supercomputing e.V. (grant CHTB00) through the John von Neumann Institute for Computing (NIC). \par

This work used the DiRAC at Durham facility managed by the Institute for Computational Cosmology on behalf of the STFC DiRAC HPC Facility (www.dirac.ac.uk). The equipment was funded by BEIS capital funding via STFC capital grants ST/P002293/1, ST/R002371/1 and ST/S002502/1, Durham University and STFC operations grant ST/R000832/1. DiRAC is part of the National e-Infrastructure. \par

\section*{Data Availability}
The target catalogue with derived properties of our sample will be available on CDS on publication. Spectral fitting results will be made available upon request. \par 

The deep radio data used in this paper is from the LoTSS Deep Fields Data Release 1 which is presented in \cite{tasse_lofar_2021}. The data is made publicly available through the LOFAR Surveys website (\href{https://lofar-surveys.org/deepfields.html}{https://lofar-surveys.org/deepfields.html}).\par

We use the SDSS Quasar catalogue from DR16 which can be accessed at \href{https://www.sdss4.org/dr17/algorithms/qso_catalog/}{https://www.sdss4.org/dr17/algorithms/qso\_catalog/} and the broad-line AGN catalogue from SDSS DR7 which can be found at \href{http://cdsarc.u-strasbg.fr/cgi-bin/ftp-index?/ftp/cats/J/ApJS/243/21}{http://cdsarc.u-strasbg.fr/cgi-bin/ftp-index?/ftp/cats/J/ApJS/243/21} . \par

The SDSS spectra used in the spectral fitting can be downloaded from \href{https://dr14.sdss.org/optical/spectrum/search}{https://dr14.sdss.org/optical/spectrum/search} by uploading the source's location or plate information.



\bibliographystyle{mnras}
\bibliography{references} 


\appendix
\section{}

\begin{figure}
    \centering
    \includegraphics[width=0.5\textwidth]{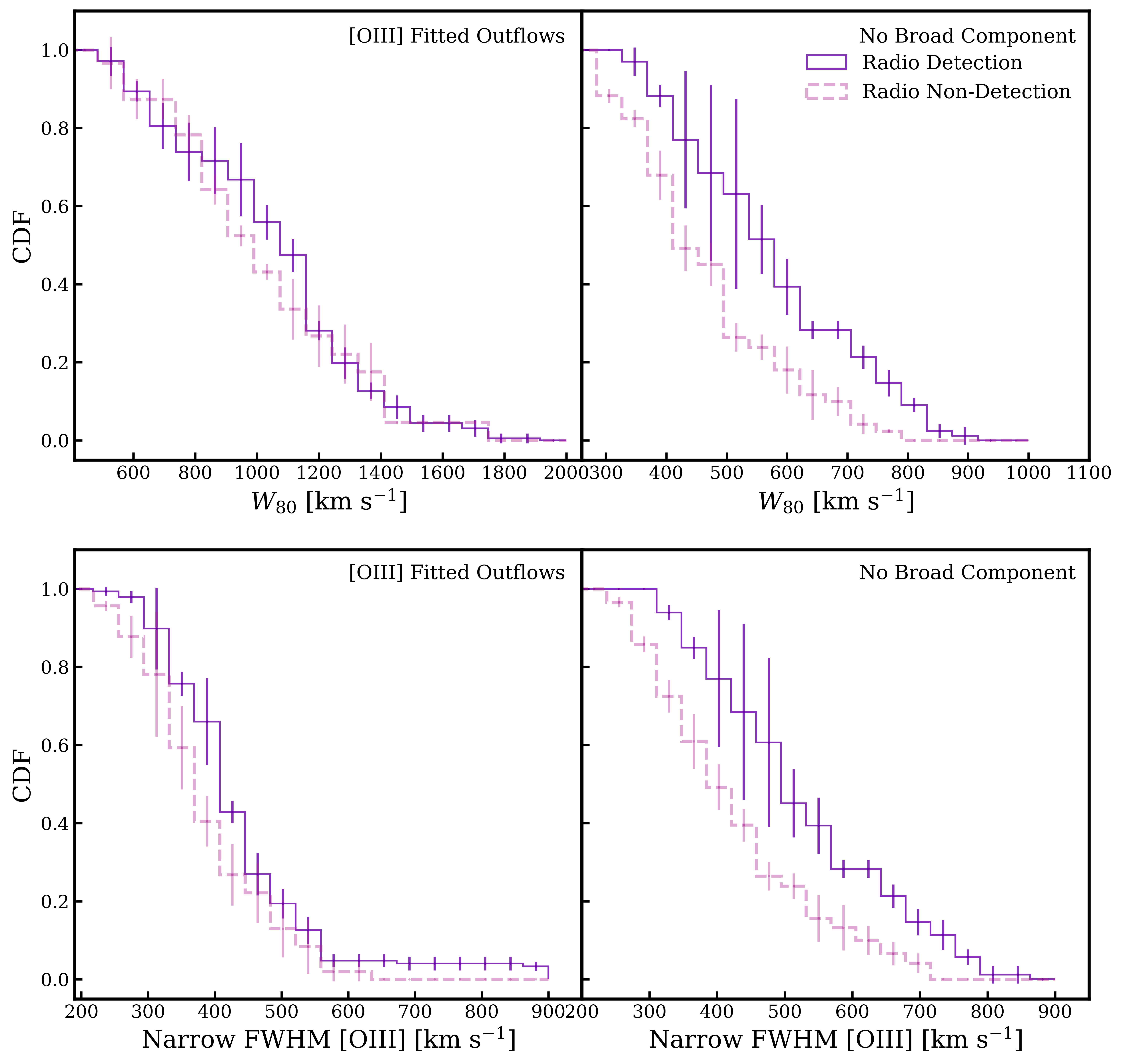}
    \caption{Further average cumulative distribution functions of [O~{\sc iii}] properties. The solid purple step function shows the distribution for the radio detected AGN and the dashed pink line shows the information from the radio non-detected sources. $\emph{Upper left:}$ CDFs of $W_{80}$ [O~{\sc iii}] showing the results only AGN which are fitted with a second component. $\emph{Upper right:}$ CDFs of $W_{80}$ [O~{\sc iii}], all AGN which are fitted with a single Gaussian. $\emph{Lower Left:}$ CDFs of the FWHM of the narrow component of [O~{\sc iii}] with  only AGN which are fitted with a second component. $\emph{Lower Right:}$ CDFs of the FWHM of the narrow component of [O~{\sc iii}] with all AGN which are fitted with a single Gaussian. The uncertainties presented are as a result of bootstrapping. }
    \label{fig:four_culul}
\end{figure}

Here we present additional CDFs to support the discussion in \ref{cumul} to further demonstrate how the [O~{\sc iii}] line profiles differ in the radio detected AGN compared to the radio non-detected AGN. The top panels of \ref{fig:four_culul} show the results for the CDFs of $W_{80}$ for AGN with [O~{\sc iii}] fitted outflows (upper left) and AGN with only a single component fitted to [O~{\sc iii}] (upper right). The lower two panels show the CDFs for the FWHM of the narrow component, with the lower left panel containing AGN with an [O~{\sc iii}] fitted outflow and the lower right panel containing AGN with no broad component, so only a narrow component is fitted to [O~{\sc iii}]. \par

\bsp	
\label{lastpage}
\end{document}